\documentclass[10pt]{article}
\usepackage{fancyhdr}
\usepackage{extramarks}
\usepackage{amsmath}
\usepackage{amsthm}
\usepackage{amsfonts}
\usepackage{siunitx}
\usepackage{tikz}
\usepackage{algorithm}
\usepackage{algpseudocode}
\usepackage{multirow}
\usepackage[export]{adjustbox}
\usepackage{booktabs}
\usepackage{palatino}
\usepackage{graphicx}
\usepackage{subfigure}

\usepackage[colorlinks,linkcolor=black,anchorcolor=black,citecolor=black,urlcolor=blue]{hyperref}
\usepackage{amsmath,bm}
\usepackage{booktabs}
\usepackage{mathtools}
\usepackage{amssymb}
\usepackage{caption}
\usepackage{capt-of}
\usepackage{mciteplus}
\usepackage{cite}
\usepackage{mathrsfs}
\usepackage[title,titletoc,toc]{appendix}
\usepackage{xr}
\usepackage{parskip}
\usepackage{soul}
\usepackage{textcomp}
\usepackage[colaction]{multicol}
\usepackage[switch]{lineno}
\usepackage{lipsum}
\usepackage{etoolbox}
\usepackage{longtable}
\usepackage{array}
\usepackage{tablefootnote}
\usepackage{ragged2e}

\usepackage{titlesec}

\newcolumntype{C}[1]{>{\centering\arraybackslash}p{#1}}
\usetikzlibrary{automata,positioning}
\usetikzlibrary{automata,positioning}

\begin{document}

\title{Multiscale Topology in Interactomic Network: From Transcriptome to Antiaddiction Drug Repurposing}
\author{Hongyan Du$^{1,2}$ and Guo-Wei Wei$^{2,3,4}$\footnote{
		Corresponding author.		Email: weig@msu.edu} and Tingjun Hou$^{1}$\footnote{
            Corresponding author.		Email: tingjunhou@zju.edu.cn} \\
$^1$ College of Pharmaceutical Sciences, \\
Zhejiang University, Hangzhou 310058, Zhejiang, China. \\
$^2$ Department of Mathematics, \\
Michigan State University, MI 48824, USA.\\
$^3$ Department of Electrical and Computer Engineering,\\
Michigan State University, MI 48824, USA. \\
$^4$ Department of Biochemistry and Molecular Biology,\\
Michigan State University, MI 48824, USA.  
}
\date{\today} 

\maketitle

\begin{abstract}
    The escalating drug addiction crisis in the United States underscores the urgent need for innovative therapeutic strategies. This study embarked on an innovative and rigorous strategy to unearth potential drug repurposing candidates for opioid and cocaine addiction treatment, bridging the gap between transcriptomic data analysis and drug discovery. We initiated our approach by conducting differential gene expression analysis on addiction-related transcriptomic data to identify key genes. We propose a novel topological differentiation  to identify key genes from a protein-protein interaction (PPI) network derived from DEGs. This method utilizes persistent Laplacians to accurately single out pivotal nodes within the network, conducting this analysis in a multiscale manner to ensure high reliability. Through rigorous literature validation, pathway analysis, and data-availability scrutiny, we identified three pivotal molecular targets, mTOR, mGluR5, and NMDAR, for drug repurposing from DrugBank. We crafted machine learning models employing two natural language processing (NLP)-based embeddings and a traditional 2D fingerprint, which demonstrated robust predictive ability in gauging binding affinities of DrugBank compounds to selected targets. Furthermore, we elucidated the interactions of promising drugs with the targets and evaluated their drug-likeness. This study delineates a multi-faceted and comprehensive analytical framework, amalgamating bioinformatics, topological data analysis and machine learning, for drug repurposing in addiction treatment, setting the stage for subsequent experimental validation. The versatility of the methods we developed allows for applications across a range of diseases and transcriptomic datasets.
 
\end{abstract}


\newpage

{\setcounter{tocdepth}{4} \tableofcontents}

\clearpage \pagebreak

\setcounter{page}{1}
\renewcommand{\thepage}{{\arabic{page}}}

\clearpage

\section{Introduction}
The ongoing drug addiction crisis presents a severe global public health challenge, particularly in the United States (US).  The startling statistics from the US National Center for Drug Abuse Statistics shows that as of 2020, 37.309 million individuals aged 12 and older were identified as current illegal drug users, with a cumulative nearly one million drug overdose deaths recorded since 2000. The economic burden is significant, with \$35 billion allocated for drug control in 2020. Moreover, the mortality associated with drug addiction is harrowing, with nearly 92,000 drug overdose deaths recorded in 2020 alone. Among the diverse substances abused, opioids and cocaine are two predominant agents exacerbating the addiction epidemic.

The term ``opioid" encompasses both natural and synthetic substances that bind to specific opioid receptors within the human body. Misuse of opioids can lead to opioid use disorder (OUD), characterized by cravings, continued use despite physical and/or psychological decline, increased tolerance, and withdrawal symptoms upon cessation \cite{mclellan2000drug}. Opioids interact primarily with three receptors in the central nervous system (CNS) and peripheral organs: the mu opioid receptor (MOR), kappa opioid receptor (KOR), and delta-opioid receptor (DOR), regulating various physiological functions including analgesia, respiration, and hormonal regulation \cite{zaki1996opioid}. Synthetic and exogenous opioids (e.g., morphine, heroin, oxycontin) mimic endorphins in their action on opioid receptors, with repeated and escalating use inducing gradual brain adaptations. 

The    United States Food and Drug Administration (FDA) has approved three medications for opioid dependency management: methadone, buprenorphine, and naltrexone. Methadone, a full MOR agonist, alleviates withdrawal and craving symptoms, proving beneficial in methadone maintenance treatment (MMT) to mitigate withdrawal severity and deter additional opioid intake for euphoria. While MMT aids in reducing cravings, facilitating treatment retention, methadone carries a risk of respiratory depression if misadministered\cite{modesto2010methadone}. Buprenorphine is a partial MOR agonist and serves as an alternative to methadone. It has a ceiling effect on MOR stimulation, offering less euphoria and a lower risk of respiratory depression\cite{mattick2014buprenorphine}. Naltrexone is an MOR antagonist that effectively reduces drug cravings and the risk of overdose without inducing sedation, analgesia, or euphoria\cite{gastfriend2011intramuscular}. However, its usage is less common due to lower patient acceptance and adherence. While current medications effectively manage OUD to an extent, relapse and remission are prevalent due to the neurobiological shifts and opioid receptor tolerance from repeated abuse. The identification of new therapeutic targets and corresponding drug development is anticipated to alleviate the challenges encountered in OUD treatment. 

Cocaine is a tropane alkaloid stimulant known for its high addictive potential. Its misuse leads to serious health complications, including an increased risk of human immunodeficiency virus (HIV), hepatitis B, and heart disease. Additionally, it is associated with a rise in crime and violence rates\cite{de2009crack}. Cocaine exerts its effects by binding to three monoamine transporters: dopamine transporter (DAT), serotonin transporter (SERT), and norepinephrine transporter (NET)\cite{beuming2008binding, kristensen2011slc6, elliott2005psychostimulants}. This binding action inhibits neurotransmitter reuptake, leading to elevated synaptic neurotransmitter levels and enhanced euphoric experiences. Specifically, cocaine's blockade of DAT increases dopamine levels in the synaptic cleft, which in turn stimulates dopamine receptors in the postsynaptic neuron, generating euphoria and arousal\cite{cheng2015insights}. The D3 dopamine receptor (D3R) in the mesolimbic reward system plays a crucial role in cocaine's rewarding and addictive effects\cite{matuskey2014dopamine}. In vivo studies reveal that acute cocaine intoxication boosts serotonin release, whereas withdrawal leads to decreased serotonin levels in the nucleus accumbens\cite{broderick2004clozapine}. NET, embedded in the plasma membrane of noradrenergic neurons, serves as a primary conduit for the inactivation of noradrenergic signaling by reabsorbing synaptically released norepinephrine (NE). Preclinical research highlights the role of the noradrenergic system in stress-induced reinstatement of cocaine seeking behavior\cite{leri2002blockade}. Some clinical studies suggest that adrenergic blockers might offer a potential treatment pathway, especially for individuals with severe cocaine withdrawal symptoms\cite{kampman2006double}. Despite ongoing research efforts, the  FDA  has not approved any effective medication for treating cocaine dependence. 

Differentially Expressed Gene (DEG) analysis is a crucial tool in molecular biology and genetics, showing promise in identifying new targets for drug addiction treatment, and aiding in discovering novel therapeutic agents. By exploring the transcriptomic dynamics underlying various biological conditions including drug addiction, DEG analysis allows researchers to examine transcriptomic datasets and identify genes with significantly altered expression levels under different conditions or between varying sample groups. Comparing gene expression profiles between drug-exposed and control groups unveils potential molecular mechanisms driving addictive behaviors or responses to substances of abuse. Zhang et al. have delineated biomarkers for opioid addiction through integrated bioinformatics analysis of gene expression data from heroin addicts\cite{zhang2020identification}. Similarly, Wang et al. employed this approach to explore cocaine addiction and suggest potential therapeutic drugs targeting key genes, as identified through database analysis\cite{wang2023identification}. Despite their valuable insights, these studies primarily rely on central algorithms for key gene identification from DEGs' PPI networks. These algorithms focus predominantly on connection relationships within the network, often neglecting the quantitative assessment of interaction confidence or strength, thereby potentially missing critical information. Furthermore, these methods are traditionally limited in handling only the low-dimensional connections of data, which may result in an incomplete understanding of the complex, high-dimensional nature of PPI networks. Moreover, despite identifying key genes, these studies did not employ quantitative methodologies to drive drug discovery efforts targeting the identified genes. 

Topological Data Analysis (TDA) introduces a groundbreaking perspective in analyzing these intricate biological networks. Within TDA, persistent homology stands out as a powerful tool,  applying algebraic topology to uncover topological features such as holes and voids \cite{zomorodian2004computing}. This technique facilitates a multiscale analysis through its filtration process, generating a series of topological invariants that uniquely characterize data.  However, persistent homology does not account for the homotopic shape evolution of data, a limitation addressed by persistent spectral graph (PST), also called persistent Laplacians \cite{wang2020persistent}. PST encompasses both harmonic and non-harmonic spectra, where the former recovers all topological invariants from persistent homology, and the latter reveals the homotopic shape evolution. It represents PPI networks in a topological space and adeptly captures both the overarching structure and subtle local interactions within the PPI network. PST has been successfully employed in various applications, including machine learning-assisted protein engineering predictions\cite{qiu2023persistent}, forecasting dominant SARS-CoV-2 variants\cite{chen2022persistent}, predicting protein-ligand binding affinity\cite{meng2021persistent}, drug addiction analysis \cite{zhu2023tidal}, and in dimensionality reduction for gene expression data\cite{cottrell2023plpca}.

Investigating existing drugs for new therapeutic prospects beyond their initially intended applications, a process known as drug repurposing, has yielded successful outcomes in several scenarios, providing a pathway to diminish development costs and expedite timelines\cite{pushpakom2019drug}. In the wake of the proliferation of biological data, machine learning has ascended as a pivotal instrument in the domain of drug discovery. Utilizing nonlinear regression, machine learning predictions harness available datasets to disclose inherent patterns. In light of the prevailing data complexity challenges, machine learning-based screening frequently surpasses physics-based methodologies like molecular docking and molecular dynamics (MD) simulations, facilitating a swift screening process of potential drug candidates within expansive chemical libraries. Feng et al. employed machine learning to repurpose compounds from DrugBank targeting MOR, KOR, and DOR, and executed an exhaustive analysis on the binding conformations and drug-likeness of potential drugs with a high predictive binding affinity\cite{feng2023machine}. 

\begin{figure}[H]
	\centering
	\includegraphics[width = 1\textwidth]{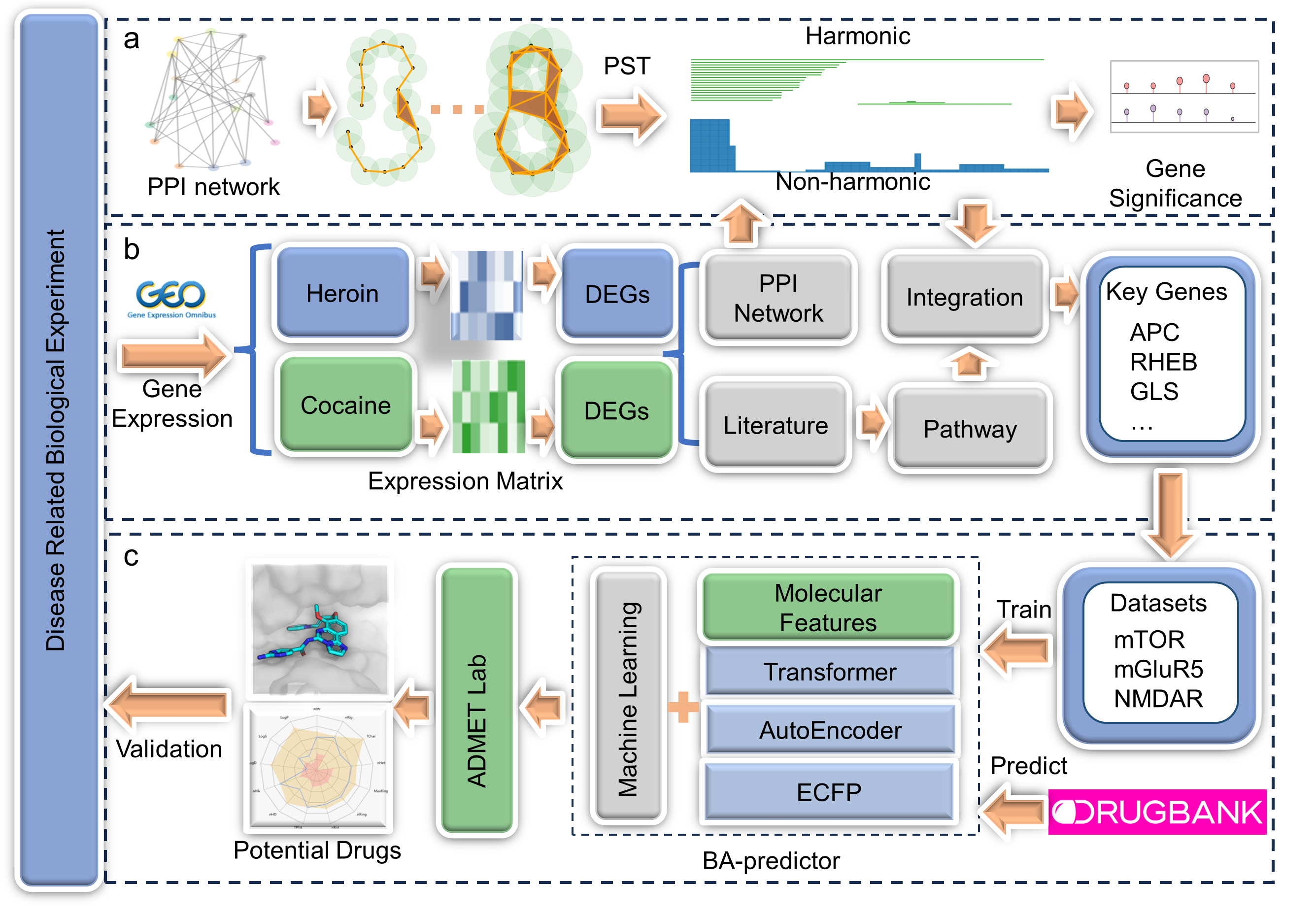}
	\caption{Overview of the Study: (a) PST-Based Topological Differentiation Analysis: This stage involves applying PST for topological differentiation analysis of the PPI network. The method quantifies the significance of nodes within the network in a multiscale manner. (b) DEG Analysis: We extracted opioid and cocaine addiction-related transcriptomic data from the GEO database. Key genes were identified from the PPI network derived from DEGs using topological analysis, and results were integrated across networks with various thresholds. Literature validation and pathway analysis were conducted to confirm the functionality and biological mechanisms of the key genes in substance addiction. (c) Drug Repurposing: Machine learning models, incorporating NLP-based fingerprints and traditional 2D fingerprints, were developed to predict the binding affinities of DrugBank compounds to three addiction-related targets: mTOR, mGluR5, and NMDAR.  This process aims to identify potential repurposing candidates after the ADMET analysis for treating substance addiction.}
	\label{fig:workflow}
\end{figure}

In this study, we present a multi-faceted and rigorous strategy that successfully bridges the complex gap between transcriptomic data analysis and drug discovery in the context of substance addiction (Figure \ref{fig:workflow}). We have  utilized PST to conduct a topological differentiation of the PPI network derived from DEG data. This innovative approach provides a powerful tool for identifying key genes implicated in opioid and cocaine addiction. The application of PST in our study brings several significant advantages: it facilitates a multiscale analysis through an intricate filtration process, quantitatively assesses the confidence or strength of each interaction, and extracts high-dimensional information from the PPI network. These features are instrumental in capturing the complex dynamics and interactions within the network, thus providing a more comprehensive understanding of the underlying biological mechanisms. Following a rigorous validation through literature review, pathway analysis, and data-availability scrutiny, we identify three targets highly pertinent to substance addiction: mTOR, mGluR5, and NMDAR for DrugBank repurposing. We devised machine learning (ML) models employing two natural language processing (NLP)-based embeddings generated via transformer and autoencoder models, alongside a traditional 2D fingerprint Extended Connectivity Fingerprint (ECFP). These models demonstrated robust predictive capacity in five-fold cross-validation tests. Leveraging these ML models, we conducted a systematic evaluation of the binding affinities of DrugBank compounds to these targets across various binding thresholds. This examination led to the identification of several drugs exhibiting satisfactory binding energies at specified binding affinity thresholds. Subsequently, molecular docking was performed on a select group of promising drugs to elucidate their interactions with receptors. Additionally, we have conducted a thorough ADMET (absorption, distribution, metabolism, excretion, and toxicity) analysis, providing a comprehensive evaluation of the pharmacokinetic and safety profiles of potential therapeutic compounds. The drugs identified, with their potent receptor inhibition affinities and favorable ADMET profiles, are prime candidates for subsequent biological experiment. This study establishes a new paradigm in drug repurposing research for addiction treatment, seamlessly integrating bioinformatics, topological data analysis, and machine learning into an inclusive and systematic analytical framework. The adaptability of this framework to a wide range of diseases and transcriptomic datasets highlights its potential to significantly advance drug discovery across multiple medical fields.

\section{Results and Discussion}

\subsection{Opioid Addiction Analysis}

\subsubsection{Differential expression analysis}
We obtained the GSE87823 dataset from the Gene Expression Omnibus (GEO) database to investigate the gene expression alterations associated with opioid addiction. This dataset encompasses expression data from the human nucleus accumbens, comparing heroin users with controls. The nucleus accumbens is pivotal in regulating reward, emotion, motivation, and goal-directed behavior, making it a focal point for studies on addiction\cite{gibson2019nucleus}. Heroin, chemically known as 3,6-diacetylmorphine or diamorphine, is a semi-synthetic morphine derivative. It was first synthesized in 1874 and later introduced commercially as a remedy for cough and certain respiratory diseases\cite{sneader1998discovery}. However, its potent analgesic effects soon overshadowed its therapeutic uses when it was discovered that heroin could induce intense dependence, leading to illicit consumption\cite{wolff1956heroin}. As a result, heroin rapidly became a significant public health concern. Today, it stands as the predominant opioid of abuse in many countries\cite{milella2023heroin}. Intriguingly, despite its widespread misuse, the psychopharmacology of heroin remains less explored than that of other drugs of abuse, such as cocaine and general psychostimulants\cite{milella2023heroin}.

\begin{figure}[H]
	\centering
	\includegraphics[width = 1\textwidth]{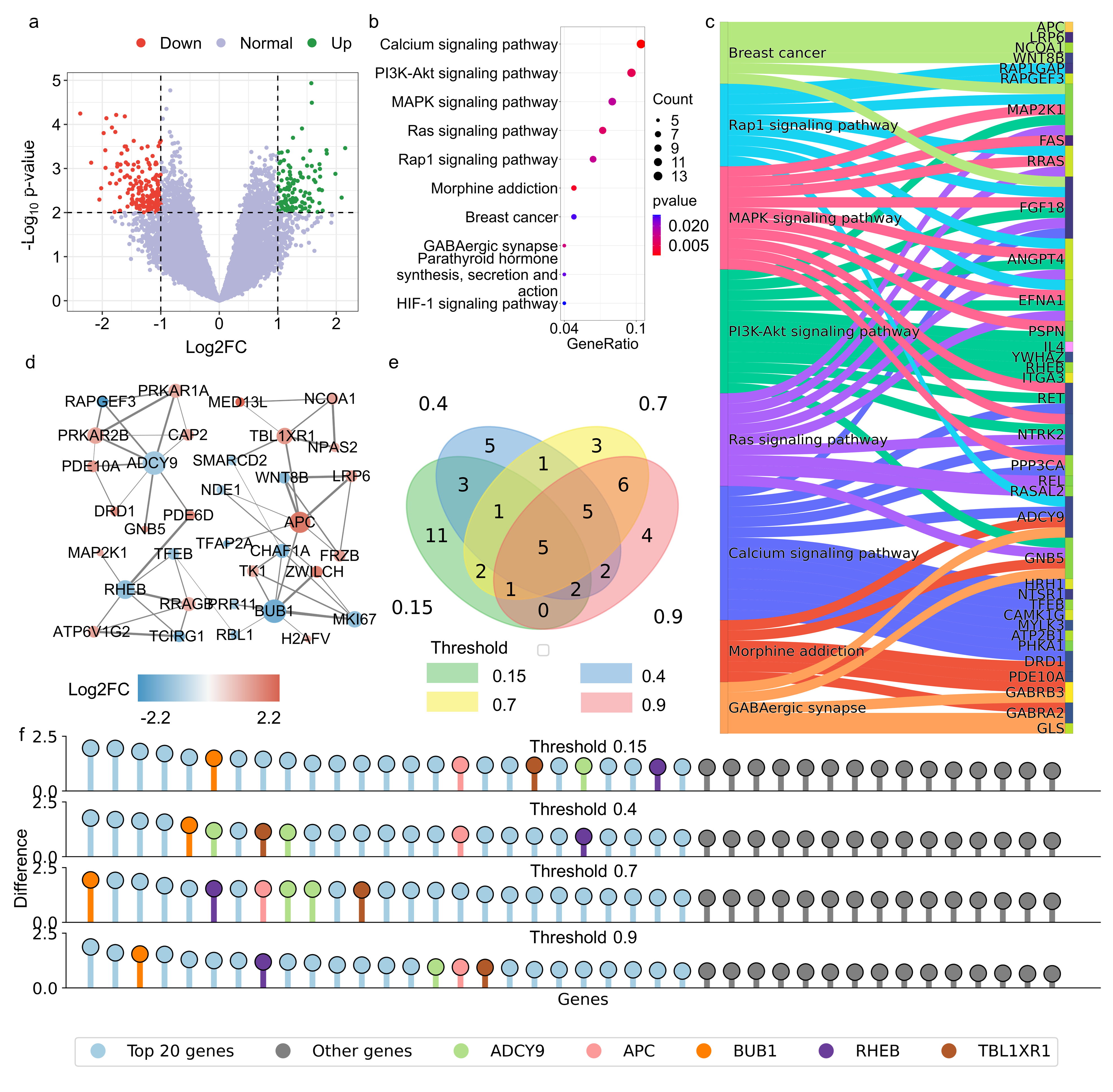}
	\caption{DEG analysis for opioid addiction. (a) Volcano plot of DEGs: This plot visually distinguishes DEGs in opioid addiction, highlighting significant genes above the threshold lines. (b) Enriched pathways in opioid addiction: Showcases the top ten pathways significantly enriched in the context of opioid addiction, emphasizing their relevance in the disease mechanism. (c) Sankey plot of pathway-DEG relationships: Illustrates the top eight enriched pathways and their connections with DEGs, highlighting the intricate interplay between them. (d) Key gene-related PPI sub-network: Depicts the PPI sub-network specifically associated with key genes identified in opioid addiction. (e) Venn diagram of significant intersections: Displays intersections of the top 25 significant genes across four PPI networks with varying thresholds, demonstrating the consistency of key genes in opioid addiction. (f) PST-based network differentiation significance: This graph presents the significance of individual genes as calculated from the PST-based differentiation of the network. For clarity, only the first 40 genes are included.}
	\label{fig:opioid_deg}
\end{figure}

We discerned a total of 295 DEGs in our study: 124 were up-regulated, while 171 were down-regulated (Figure \ref{fig:opioid_deg}a). This distinction provides us with an initial glimpse into the potential molecular mechanisms underlying opioid addiction, as genes that are differentially expressed can hint at pathways and processes that are altered in the disease state. 

\subsubsection{Multiscale topological differentiation of PPI networks}

To further investigate    the potential functional interactions among these DEGs, and to pinpoint those that may act as hubs or central players, we retrieved the PPI network for these DEGs from the STRING database (Figure \ref{fig:opioid_deg}b). STRING, a comprehensive database, provides known and predicted protein associations, derived from computational prediction, knowledge transfer between organisms, and interactions aggregated from other databases\cite{szklarczyk2023string}. By establishing various interaction confidence thresholds (0.15, 0.4, 0.7, and 0.9), we facilitated a multi-resolution analysis. These thresholds enable us to identify robust interactions while also allowing for potential weaker, yet biologically relevant, interactions to be captured. 

\begin{figure}[H]
	\centering
	\includegraphics[width = 0.7\textwidth]{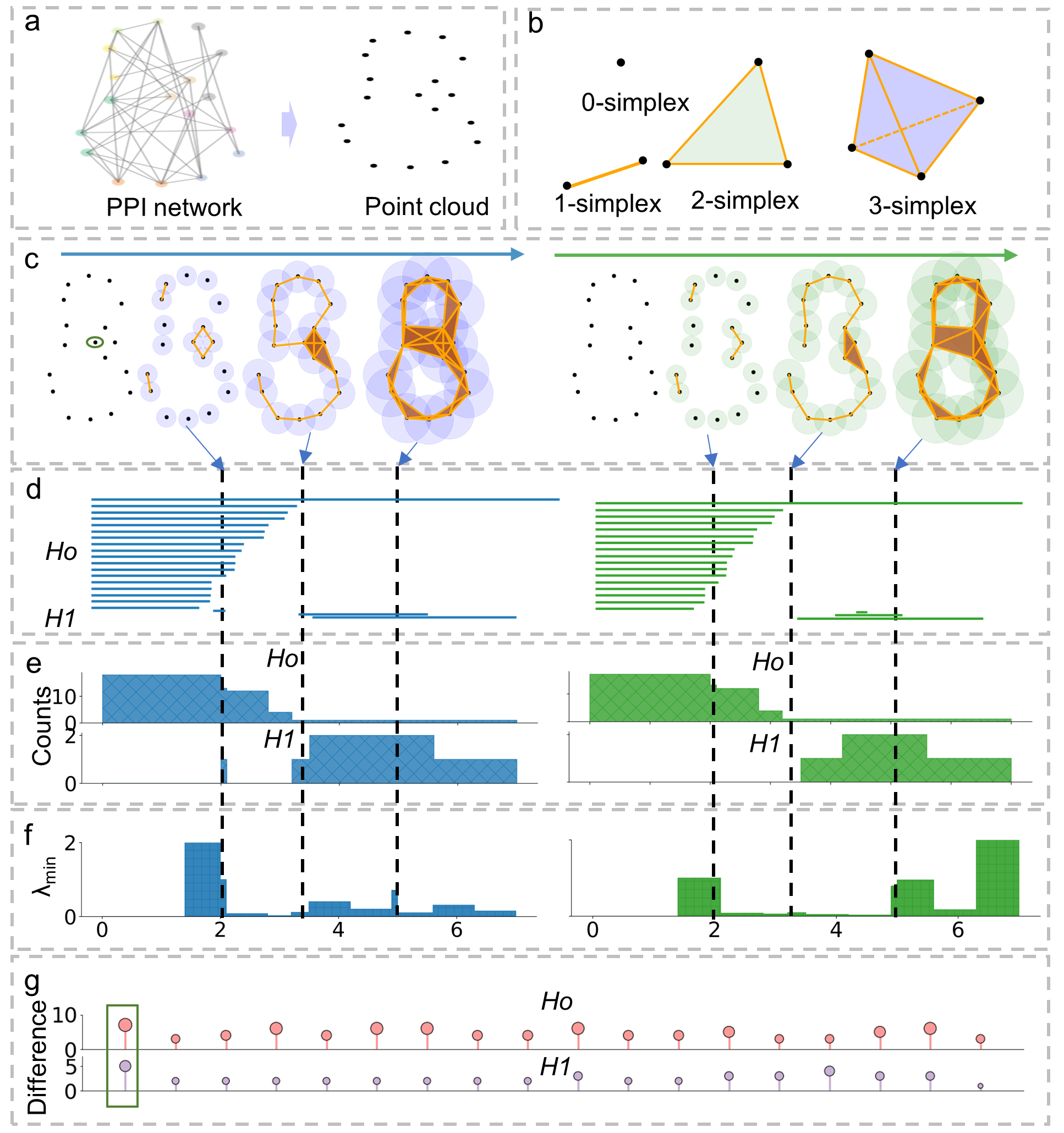}
	\caption{Topological differentiation of network. (a) PPI network as a point cloud: This panel visualizes the PPI network abstracted as a point cloud, forming the basis of a simplicial complex. (b) Basic unit of simplicial complex. (c) Filtration process: This panel depicts the filtration process, which generates a series of simplicial complexes with increasing radii. The left figure shows the original PPI network's simplicial complex, while the right figure displays the new simplicial complex formed after deleting a protein (indicated by the green circle). PH and PST are used to characterize topological and geometric changes post-deletion. (d) Persistent barcodes in topological representation: PH is utilized here to provide a topological representation of the network, illustrated through persistent barcodes. (e) PST is applied to analyze the spectra of persistent Laplacians, with harmonic spectra indicating topological persistence, akin to PH. The figure shows changes in the count of topological invariants during filtration. (f) Capturing homotopic shape Evolution: The non-harmonic spectra in PST highlight the homotopic shape evolution of data. This panel demonstrates the change in the minimum of non-harmonic spectra during the filtration process. (g) Impact of node deletion on topological invariants: This figure illustrates the changes in topological invariants resulting from the deletion of each node in the network. The sum of changes during filtration is shown, with the green rectangle highlighting the changes corresponding to the most significant node.}
	\label{fig:topological_concept}
\end{figure}

At the core of our analytical framework is the novel method we termed ``multiscale topological differentiation of networks." As depicted in Figure \ref{fig:topological_concept}, this method can leverage persistent spectral theory and persistent homology to provide a multidimensional analysis of the network's topological and geometric characteristics. By constructing a simplicial complex through a carefully designed filtration process, our approach vividly captures the dynamic interplay of protein interactions across various scales.

A key aspect of our methodology is the quantitative nature of the filtration process, based on the distance between proteins, derived from the confidence scores of each interaction pair. This precise and measurable analysis is a fundamental departure from conventional methods predominantly based on centrality algorithms, which often overlook such quantitative interaction information. Our technique exemplifies the application of TDA, effectively tracking the evolution of topological invariants while accounting for changes in homotopic shapes. Our methodology is characterized by its dynamic exploration of the network structure. By systematically removing specific proteins and observing the resulting topological and geometric shifts, we assess the network's structural robustness and the critical role of individual proteins. This analysis allows us to identify key proteins whose modification significantly impacts the network's overall structure.

In this study, we employ PST to meticulously analyze the spectra of persistent Laplacians for each network. This analysis involves a detailed extraction of topological persistence from the harmonic spectra, complemented by insights into the homotopic shape evolution gleaned from the non-harmonic spectra. We then vectorize these extracted features, creating a comprehensive representation of the network's topological characteristics. We calculate the Euclidean distance between vectorized features before and after the deletion of a protein. This distance serves as a metric for assessing the structural impact of the protein's absence, thereby quantifying the protein's importance within the network. Further enhancing our method's precision, we rank the nodes in each network based on their importance under various threshold settings. By identifying and intersecting the top-ranked protein nodes across all four threshold-defined networks, we pinpoint key genes. This intersection approach ensures that the genes we classify as `key' are consistently influential across multiple network scales, highlighting their potential critical role in the substance addiction related biological processes under study. Applying this methodology, we have successfully identified six key genes that hold significant importance across the networks: APC, BUB1, ADCY9, RHEB and TBL1XR1 (Figure \ref{fig:opioid_deg}e and \ref{fig:opioid_deg}f). Their consistent presence across different network thresholds underscores their potential central role in addiction-related biological pathways. 

\subsubsection{Literature validation}

To gain deeper insights into the intricate relationship between our identified key genes and opioid addiction, we turned to literature validation. Our findings emphasize the significant roles of RHEB, ADCY9, APC, and TBL1XR1 in opioid addiction pathways. RHEB interacts with the mTORC1 pathway, implicating its role in opioid tolerance and hyperalgesia, and suggesting mTOR inhibitors as potential therapeutic agents. ADCY9, pivotal in the cAMP-PKA-GABA pathway, affects neuronal excitability and dopamine release, crucial in addiction mechanisms. APC, through its involvement in the Wnt pathway, is linked to opioid withdrawal symptoms and hyperalgesia. Lastly, TBL1XR1, while less explored, shows potential relevance to opioid addiction via the Wnt pathway. These genes illustrate the multifaceted molecular dynamics of opioid addiction. For more detailed information and literature references, please refer to the supporting information.

\subsubsection{Pathway enrichment}
To gain a deeper insight into the broader biological processes influenced by the DEGs and to pinpoint significant biological pathways, we conducted pathway enrichment analysis. Figure \ref{fig:opioid_deg}c and \ref{fig:opioid_deg}d depict the top ten enriched signaling pathways. Notably, the three most prominent pathways are the Calcium signaling, PI3K-Akt pathway, and MAPK pathway, involving 13, 12, and 10 DEGs, respectively. Notably, six DEGs have been identified in Morphine addiction pathway. 

The intricate link between opioid addiction and calcium signaling has been extensively researched. Upon ligand binding to the opioid receptor, there is a resultant dissociation of the $\alpha $-GTP complex from the $\gamma $ dimer subunits. This $\gamma $ dimer subsequently acts to directly inhibit the calcium channels, leading to a reduction in the intracellular calcium concentration\cite{moises1994mu}. The modulatory effect of opioid receptor activation on calcium channel activity has been corroborated across various brain regions, including the hippocampus, nucleus locus coeruleus, and the area postrema, among others\cite{zamponi2002modulating}. K\"{o}nig et al. found that PI3K$\gamma $ modulates the desensitization of the mu opioid receptor\cite{konig2010modulation}. Madishetti et al. discovered the essential function of PI3K$\gamma $ in cAMP-mediated inflammatory hypernociception\cite{madishetti2014pi3kgamma}. The Mitogen-Activated Protein Kinase (MAPK) pathway is closely associated with opioid addiction. Several genes within MAPK pathway have been identified in relation to opioid receptor signaling and behavior, including ERK 1/2\cite{miyatake2009inhibition}, c-Jun N-terminal kinase\cite{shahabi2006delta} and p38 MAPK\cite{tan2009p38}. 

\subsection{Cocaine Addiction Analysis}

\subsubsection{Differential expression analysis and key genes identification}
\begin{figure}[H]
	\centering
	\includegraphics[width = 1\textwidth]{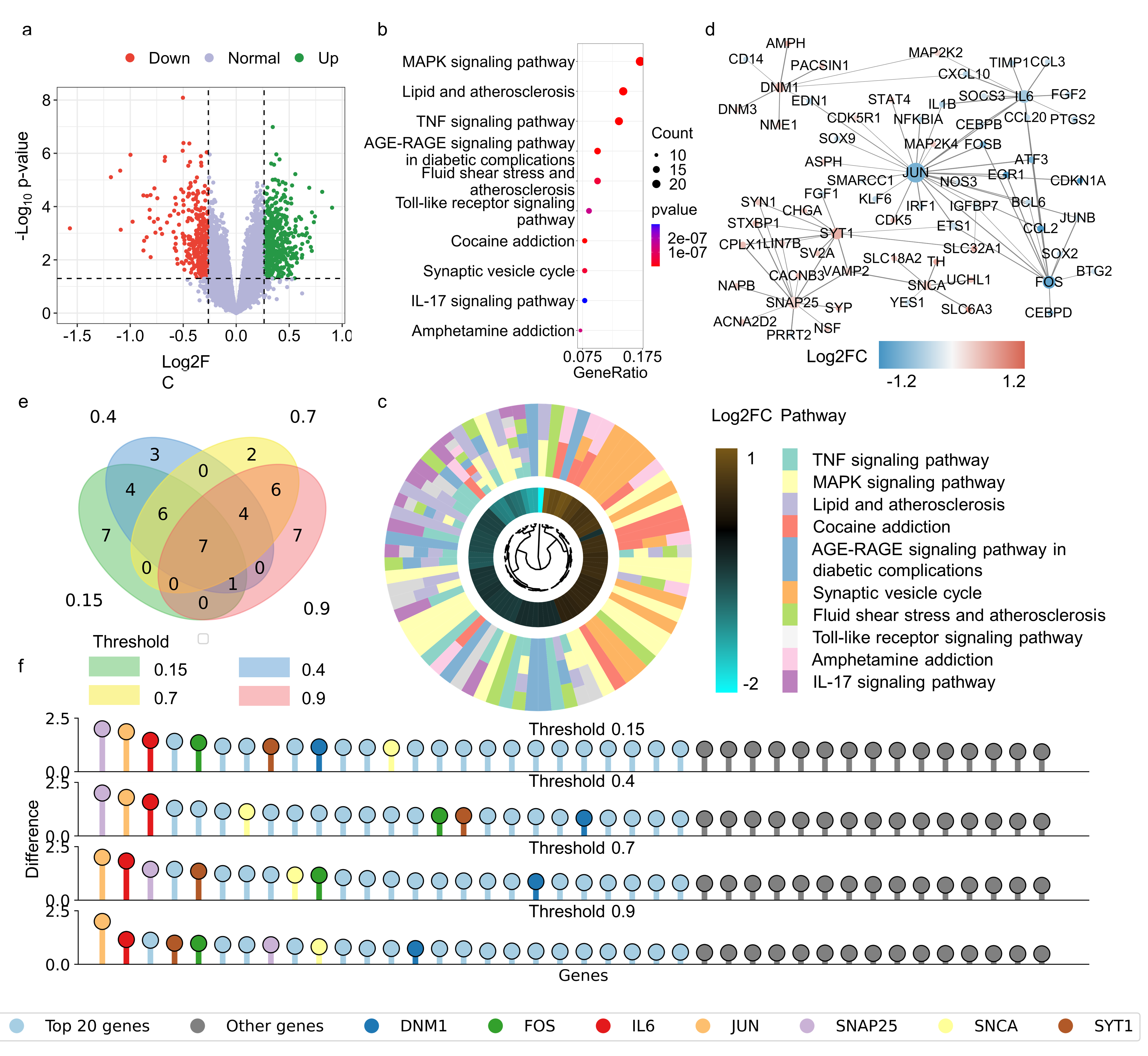}
	\caption{DEG analysis for cocaine addiction. (a) Volcano plot of DEGs: This plot visually distinguishes DEGs in cocaine addiction, highlighting significant genes above the threshold lines. (b,c) Enriched pathways in cocaine addiction: Showcases the top ten pathways significantly enriched in the context of cocaine addiction, emphasizing their relevance in the disease mechanism. (d) Key gene-related PPI sub-network: Depicts the PPI sub-network specifically associated with key genes identified in cocaine addiction. (e) Venn diagram of significant intersections: Displays intersections of the top 25 significant genes across four PPI networks with varying thresholds, demonstrating the consistency of key genes in cocaine addiction. (f) PST-based network differentiation significance: This graph presents the significance of individual genes as calculated from the PST-based differentiation of the network. For clarity, only the first 40 genes are included.}
	\label{fig:cocaine_deg}
\end{figure}

We procured and analyzed the GSE54839 dataset from the GEO database to scrutinize the alterations in gene expression consequential to cocaine addiction\cite{bannon2014molecular}. This dataset incorporates expression data derived from the human midbrain of chronic cocaine users ($n = 10$) juxtaposed against well-matched drug-free controls ($n = 10$). The midbrain, particularly the dopamine (DA)-synthesizing neurons therein, stands central to our study due to its pivotal role in addiction pathways. These neurons are largely implicated in mediating reward and pleasure centers in the brain, thereby making it crucial to discern the molecular perturbations induced by drug use, specifically in this neural locale. 
We discerned a total of 824 DEGs. Breaking down this data further revealed 467 up-regulated genes, indicating potential hyperactivity or over-expression in certain pathways, while 356 genes were observed to be down-regulated, signifying possible suppression or reduced activity (Figure \ref{fig:cocaine_deg}a). To decipher the potential interplay between these DEGs, we consulted the STRING database to draw their PPI network (Figure \ref{fig:cocaine_deg}d). Borrowing from our methodology deployed in the opioid addiction analysis, we leveraged PST to do topological differentiation analysis to identify key genes. These tools, with their distinct mechanisms, enabled us to spotlight pivotal genes within the network. Our differentiation analysis, which spanned multiple confidence thresholds, converged on seven key genes: DNM1, FOS, IL6, JUN, SNAP25, SNCA and SYT1 (Figure \ref{fig:cocaine_deg}e and \ref{fig:cocaine_deg}f).

\subsubsection{Literature validation}

To unravel the intricate interplay between key genes and cocaine addiction, we conducted literature validation. FOS (c-Fos), a significant gene in cocaine addiction, influences neuroplasticity and behavioral responses. Studies reveal that Fos's absence alters the expression of critical neuroadaptation markers and reduces behavioral sensitization to cocaine. Interleukin-6 (IL-6) shows a potential protective effect against cocaine-induced reactions, though evidence remains limited, indicating a need for further research. SYT1 has been linked to cognitive performance in cocaine addiction, with significant associations found between SYT1-rs2251214 and cocaine use disorder (CUD) vulnerability and severity. Lastly, $\alpha$-synuclein (SNCA) is recognized for its role in dopaminergic transmission and addictive behaviors. Chronic cocaine abuse leads to an upregulation of SNCA, with implications extending to other substances like alcohol. This validation underscores the complexity of genetic influences in cocaine addiction. For more comprehensive details and references, please refer to the supporting information.

\subsubsection{Pathway enrichment}

In an effort to elucidate the underlying biological processes linked to cocaine addiction, we undertook a comprehensive pathway enrichment analysis. The results, as illustrated in Figure \ref{fig:cocaine_deg}b and \ref{fig:cocaine_deg}c, highlight the ten most significant signaling pathways associated with this condition. Notably, the Cocaine and Amphetamine addiction pathways emerge as direct correlates to substance dependency. At the forefront of these pathways are the MAPK signaling cascade, the Lipid metabolism and atherosclerosis pathway, and the TNF signaling axis. It is of particular significance that the MAPK signaling cascade, previously implicated in opioid addiction studies, reemerges here, accentuating its pivotal role in the broader context of substance addiction\cite{bingor2021potentiated, sun2016molecular}. 

The association of cocaine with the Lipid metabolism and atherosclerosis pathway underscores the drug's deleterious cardiovascular implications. Many studies have documented the profound vascular influence of cocaine intake, emphasizing inflammation and atherosclerosis as predominant systemic outcomes with both acute and chronic manifestations\cite{bachi2017vascular, schwartz2010cardiovascular}. Not much literature evidence of relations between TNF signaling pathway and cocaine addiction was found. The only exception was due to Lewitus et al. who demonstrated that microglial TNF-$\alpha$  can modulate cocaine-induced neural plasticity and behavioral sensitization\cite{lewitus2016microglial}. 

\subsection{Integrated Analysis of Opioid and Cocaine Addiction DEGs}

\begin{figure}[H]
	\centering
	\includegraphics[width = 1\textwidth]{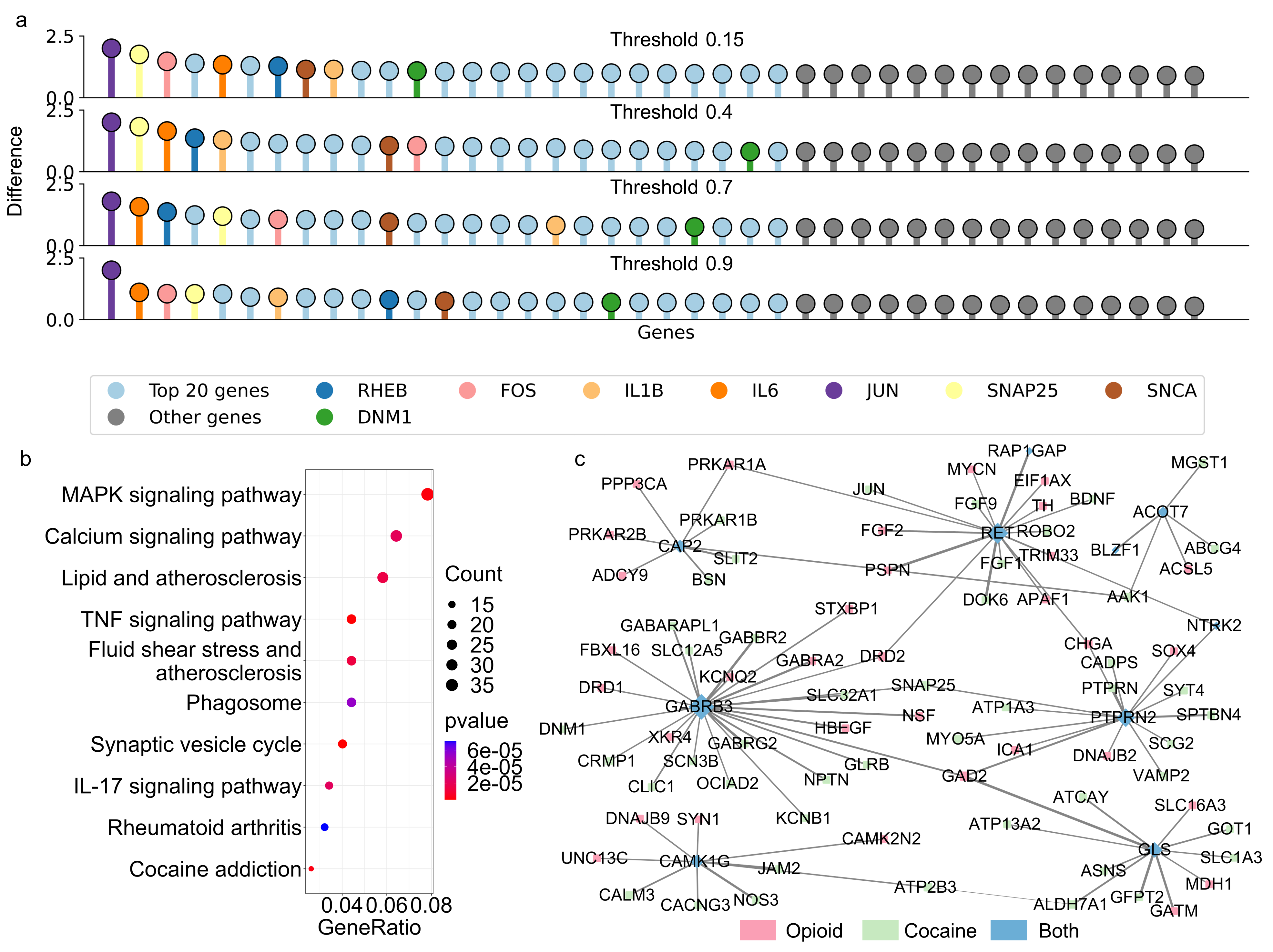}
	\caption{Integrated analysis of opioid and cocaine addiction DEGs. (a) PST-based network differentiation significance: This graph displays the significance of individual genes as determined by the PST-based differentiation in the integrated network of opioid and cocaine addiction DEGs. To enhance clarity, only the first 40 genes are shown. (b) Enriched pathways in integrated DEGs: Illustrates the pathways that are significantly enriched in the context of the integrated DEGs from both opioid and cocaine addiction, highlighting their combined biological relevance. (c) PPI sub-network related to common DEGs: Depicts the PPI sub-network associated with DEGs that are present in both opioid and cocaine addiction conditions, emphasizing the shared molecular mechanism.}
	\label{fig:both_deg}
\end{figure}

In our endeavor to unravel the shared molecular mechanisms underlying opioid and cocaine addiction, we integrated the DEGs identified in both conditions. Our primary goal was to pinpoint key genes that are consistently involved in both types of drug addictions. These genes could serve as potential therapeutic targets or biomarkers, crucial for the treatment and diagnosis of addiction. Utilizing our sophisticated topological differentiation analysis, we were able to identify eight central genes that play significant roles across both addiction types: RHEB, DNM1, FOS, IL1B, IL6, JUN, SNAP25, and SNCA (\ref{fig:both_deg}a). Notably, IL1B emerged as a novel target, a gene not previously associated with either opioid or cocaine addiction in existing analyses. In contrast, RHEB was previously recognized in the opioid addiction DEG PPI network, underscoring its importance in opioid-related mechanisms. The remaining six genes - DNM1, FOS, IL6, JUN, SNAP25, and SNCA - were identified through our analysis of cocaine addiction. Consistent with earlier studies, we conducted a pathway enrichment analysis on the integrated genes derived from both opioid and cocaine addiction analyses. As anticipated, the MAPK signaling pathway emerged as a top-ranking pathway, underscoring its crucial role in substance addiction. Figure \ref{fig:cocaine_deg}b presents the top ten pathways identified, highlighting the prominence of the MAPK pathway among them.

We then adopted a more streamlined strategy, focusing explicitly on genes that were consistently differentially expressed in both opioid and cocaine addictions. This approach revealed ten genes: CAMK1G, GABRB3, ACOT7, CAP2, PTPRN2, RET, GLS, NTRK2, BLZF1, and RAP1GAP. Most notably, seven out of these ten genes (CAMK1G, GABRB3, ACOT7, CAP2, PTPRN2, RET, and GLS) exhibited parallel expression deviations in both addiction conditions. This consistent pattern accentuates the potential significance of these genes in the broader context of drug addiction and may suggest a shared molecular pathway influencing the pathophysiology of both opioid and cocaine addictions Figure (\ref{fig:both_deg}c).

\subsubsection{Literature validation}
Extensive literature validation reveals a significant association between GABRB3, PTPRN2, GLS, and IL1B genes and drug addiction. GABRB3, encoding the $\beta$3 subunit of the GABAA receptor, is implicated in heroin dependence and alcoholism, with studies suggesting its key role in addiction pathogenesis. PTPRN2, linked to substance dependence and cognitive behavior, shows relevance in cocaine dependence and depression comorbidity, and its knockout mice models indicate altered neurotransmitter concentrations. Glutaminase (GLS), crucial in glutamate production, emerges as a central enzyme in drug reward mechanisms and relapse propensity due to its regulation of excitatory neurotransmission. Finally, IL1B, associated with inflammatory responses to cocaine, shows genetic links to alcohol and opioid dependence, highlighting its potential as a biomarker or therapeutic target in addiction. These findings illustrate the complex genetic underpinnings of addiction and underscore the cross-substance implications of these genes. For a detailed exploration of these associations and references, please refer to the supporting information.

\subsection{Repurposing of DrugBank for addiction-related targets}

\subsubsection{Binding affinity predictors for addiction-related targets}

To advance the therapeutic treatment of drug addiction and unearth potential therapeutic agents, we employed machine learning techniques to repurpose drugs from DrugBank. Machine learning models, given their ability to discern patterns and relationships in vast datasets, can effectively predict how various drugs might interact with specific biological targets associated with addiction. However, the success of these models hinges on access to a rich dataset of compound activities. Our initial search in the ChEMBL database for compounds related to our literature-validated key genes revealed a data insufficiency, precluding effective model development. This challenge redirected our focus to proteins that function downstream of these key genes. From our investigation, three proteins with substantial inhibitor data were identified: mTOR, NMDAR, and mGluR5. As we had previously deduced, RHEB's association with opioid addiction likely operates via the mTORC1 pathway. With respect to GLS, the impact on neuronal excitability from its product, glutamate, is majorly steered by NMDAR and mGluR5. Their profound connection to substance addiction has been corroborated by numerous studies\cite{corbett2023mglu5, glass2011opioid, kato2006implication, mihov2016negative}. 

We identified 4,392 mTOR inhibitors, 1,777 for mGlu5, and 2,342 for NMDAR, all characterized by either $\text{IC}_{50}$ or $K_i$ data in ChEMBL. Given the intricate nature of molecular structures, accurately representing them for machine learning models is a non-trivial task. We adopted a multi-pronged approach to molecular representation, employing both deep learning and traditional molecular fingerprint techniques. Specifically, we utilized transformer-based architectures, sequence-to-sequence autoencoders, and the established ECFP to capture the nuanced structural features of these molecules. These representations were then used to train Gradient Boosted Decision Trees (GBDT), a robust machine learning algorithm, to predict the inhibitory activities of the molecules on our target proteins. In order to offer a more robust and reliable prediction by capitalizing on the strengths of individual models, we constructed a consensus model for each target. Our model yielded Pearson correlation coefficients (R) of 0.876, 0.754, and 0.799 for mTOR, mGlu5, and NMDAR, respectively. In terms of prediction accuracy, the root mean square errors (RMSE) were found to be 0.758, 0.882, and 0.991, respectively. 

\subsubsection{Potential inhibitors of addiction-related targets in DrugBank}

To discover potential inhibitors targeting addiction-related proteins, we deployed our machine learning predictors to gauge the binding affinity of various small molecules housed in the DrugBank database. DrugBank classifies small molecules into diverse states based on their clinical statuses. For the scope of our research, we narrowed our focus exclusively on two key categories: `approved' and `investigational'. These categories are of particular interest as they encompass molecules that have either received regulatory approval or are currently under investigation in clinical studies, implying a higher likelihood of therapeutic potential and translational relevance in the context of addiction. In our analysis, we specifically selected molecules exhibiting a binding affinity greater than -9.54 kcal/mol (equivalent to a $K_i$ value of 100 nM). Such a stringent criterion, widely recognized in drug discovery, ensures that we consider only those compounds with a significant potential for therapeutic action\cite{flower2002drug}. 

\paragraph{Approved drugs with predicted efficacy on mTOR}

We embarked on an evaluation of approved drugs that demonstrated strong binding affinity towards mTOR, with our selected affinity thresholds set at -11 kcal/mol, -10 kcal/mol, and -9.54 kcal/mol. In accordance with these thresholds, we identified 4, 7, and 7 drugs, respectively, that matched our criteria (Table \ref{tab: mtor_drug}).
\begin{table}[H] 
    \centering
    \caption{Summary of the FDA-approved drugs that are potential potent inhibitors of mTOR with binding affinity (BA) < -9.54 kcal/mol.}
    \label{tab: mtor_drug}
    {
    \begin{tabular}{ ccc } 
    \toprule
    DrugID	&Name&	Predicted BA (kcal/mol)\\
    \midrule 
    DB00877	&Sirolimus&	-12.46 \\
    DB00864	&Tacrolimus&	-12.32 \\
    DB01590	&Everolimus&	-12.07 \\
    DB00337	&Pimecrolimus&	-11.39 \\
    \midrule
    DB12483	&Copanlisib&	-10.40 \\
    DB11943	&Delafloxacin&	-10.31 \\
    DB09272	&Eluxadoline&	-10.31 \\
    DB01764	&Dalfopristin&	-10.12 \\
    DB00705	&Delavirdine&	-10.07 \\
    DB00709	&Lamivudine&	-10.05 \\
    DB00615	&Rifabutin&	-10.04 \\
    \midrule
    DB00879	&Emtricitabine&	-9.92 \\
    DB00210	&Adapalene&	-9.91 \\
    DB12153	&Citicoline&	-9.87 \\
    DB12767	&Gaxilose&	-9.84 \\
    DB13274	&Micronomicin&	-9.72 \\
    DB06725	&Lornoxicam&	-9.61 \\
    DB08907	&Canagliflozin&	-9.59 \\
    \bottomrule
    \end{tabular} 
    }
\end{table}

Of particular note, both Sirolimus and Everolimus are known mTOR inhibitors\cite{klawitter2015everolimus, mackeigan2015differentiating}. Sirolimus, also known as Rapamycin, was the first pharmacological agent developed in the class of mTOR inhibitors. It received FDA approval in 1999 to deter transplant rejection in kidney recipients. It is synthesized by the bacterium Streptomyces hygroscopicus and is characterized as a macrocyclic lactone antibiotic. Apart from its primary indication, it is also utilized to manage lymphangioleiomyomatosis. In the US, a specific formulation of Sirolimus, albumin-bound Sirolimus for intravenous administration, has been indicated for treating adult patients with advanced unresectable or metastatic malignant perivascular epithelioid cell tumors (PEComa). On the other hand, Everolimus shares a functional similarity with Sirolimus in inhibiting mTOR but possesses a more favorable pharmacokinetic profile, featuring greater bioavailability, a more rapid terminal half-life, and distinct blood metabolite patterns. The FDA has approved Everolimus to prevent rejection after solid organ transplants and for treating a range of tumors, including breast, renal, and neuroendocrine tumors. Tacrolimus and Pimecrolimus are both calcineurin inhibitors, and, similar to sirolimus, they possess a macrocyclic lactone structure. Tacrolimus is primarily utilized as an immunosuppressant following kidney transplantation and is also indicated for treating atopic dermatitis when applied topically. Conversely, Pimecrolimus is specifically formulated as a topical agent for the management of mild to moderate atopic dermatitis (eczema) in patients who either do not respond well to or cannot tolerate other treatments, such as topical steroids. 

Copanlisib is prescribed for adults with recurrent follicular lymphoma, particularly those who have already received a minimum of two prior treatments. It is categorized as a phosphoinositide 3-kinase (PI3K) inhibitor, exhibiting selective inhibition towards the alpha and delta isoforms of PI3K found in malignant B-cells\cite{mensah2018spotlight}. Delafloxacin is a fluoroquinolone antibiotic used to treat specific bacterial infections in adults, such as skin infections and some pneumonias\cite{bassetti2018delafloxacin, scott2020delafloxacin}. It works by eliminating bacteria or inhibiting their growth, and is indicated for acute bacterial skin and skin structure infections (ABSSSI) and community-acquired bacterial pneumonia (CABP) caused by susceptible bacteria. Eluxadoline is prescribed to treat irritable bowel syndrome with diarrhea (IBS-D) in adults\cite{lembo2016eluxadoline}. It acts as an agonist on mu and kappa opioid receptors and an antagonist on delta opioid receptors. This modulation reduces bowel activity, stabilizes intestinal contractility, and balances stress-induced acceleration in the upper GI tract, alleviating IBS-D symptoms. 

\begin{figure}[H]
	\centering
	\includegraphics[width = 1\textwidth]{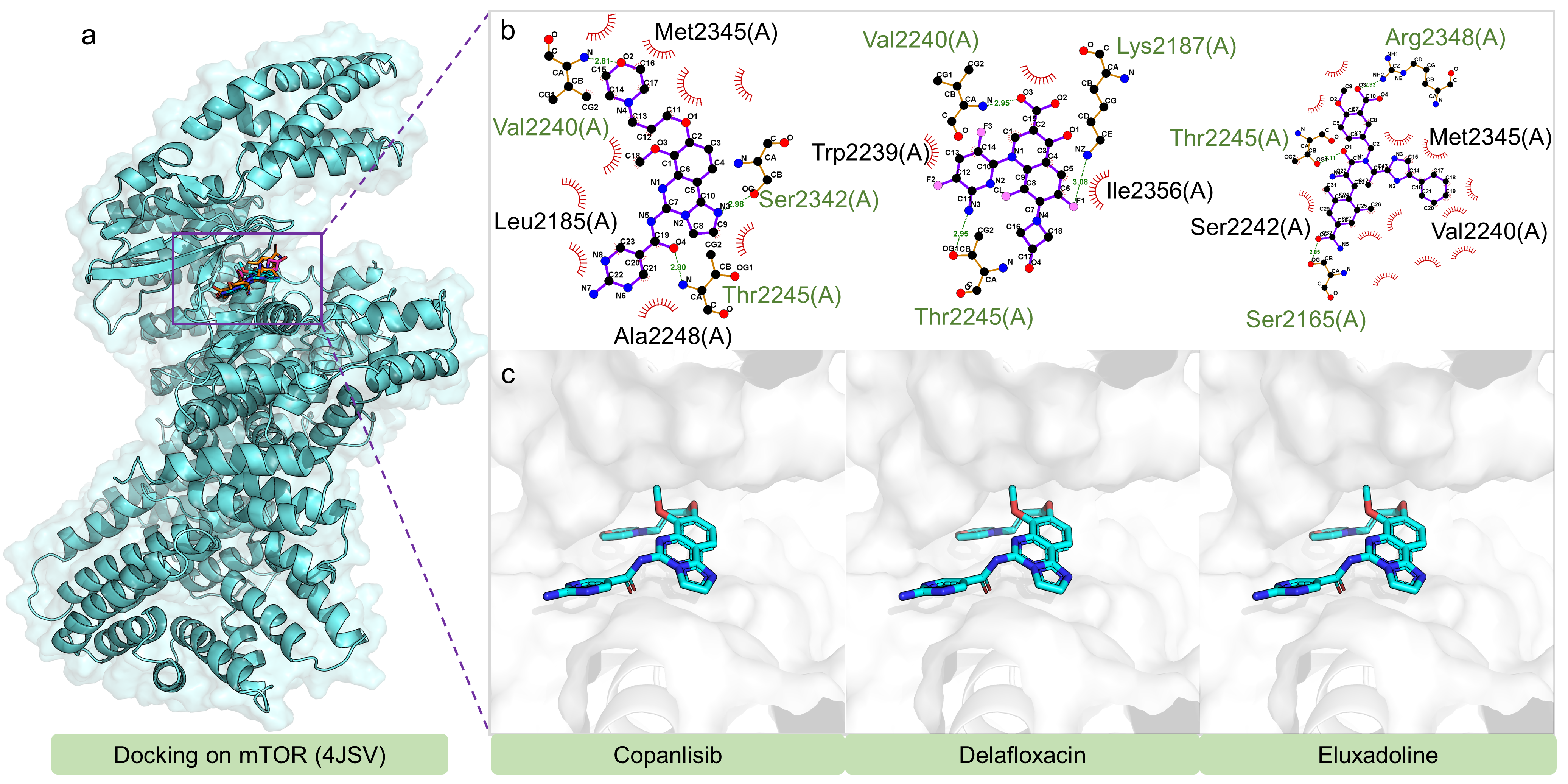}
	\caption{The docking structures and interactions of Copanlisib, Delafloxacin, and Eluxadoline with mTOR.}
	\label{fig:mtor_dock}
\end{figure}

To gain a more detailed insight into how these drugs might exert inhibitory effects on mTOR, we embarked on molecular docking analysis. This approach aimed to unveil the nuances of their binding modes and their specific interactions with the mTOR protein. As illustrated in Figure \ref{fig:mtor_dock}, Copanlisib, Delafloxacin, and Eluxadoline exhibit distinctive yet overlapping binding patterns to mTOR. One of the common characteristics among these drugs is their interaction with Thr2245. All three molecules establish hydrogen bonds with this residue: Copanlisib and Eluxadoline leverage their oxygen atoms for this interaction, while Delafloxacin utilizes its nitrogen atom. Additionally, both Copanlisib and Delafloxacin form hydrogen bonds with Val2240, but they achieve this interaction via their respective oxygen atoms. Digging deeper into their unique interactions, Copanlisib forms an additional hydrogen bond through its nitrogen atom, targeting the side chain hydroxyl group of Ser2342. Conversely, Delafloxacin establishes a connection with the side chain amino group of Lys2187, facilitated by one of its fluorine atoms. On the other hand, Eluxadoline is distinctive in its ability to form hydrogen bonds with both Arg2348 and Ser2165 through its oxygen atoms.

\paragraph{Investigational drugs with predicted efficacy on mTOR}

In our analysis of investigative drugs for potential efficacy on mTOR, we observed a greater proportion of these drugs exhibiting high affinity compared to approved drugs, as listed in Table \ref{tab: mtor_investigational}, which includes only compounds with a binding affinity better than -11 kcal/mol. Notably, while Ridaforolimus specifically targets mTOR\cite{dancey2011ridaforolimus}, the remaining compounds are dual inhibitors of PI3K and mTOR. Omipalisib (GSK2126458 or GSK458) has been investigated for treating various cancers, including acute myeloid leukemia (AML), demonstrating promising anti-leukemia effects\cite{munster2016first}. Gedatolisib (PF-05212384 or PKI-587), originally developed by Wyeth, has advanced to clinical trials for diverse cancer types\cite{shapiro2015first, wainberg2017multi}, earning FDA fast track designation for certain breast cancer treatments. PKI-179, derived from Gedatolisib, is a dual PI3K and mTOR inhibitor developed primarily for advanced malignant solid tumors\cite{venkatesan2010pki}. 
\begin{table}[H] 
    \centering
    \caption{Summary of investigational drugs that have the potential to inhibit mTOR.}
    \label{tab: mtor_investigational}
    {
    \begin{tabular}{ ccc } 
    \toprule
    DrugID	&Name&	Predicted BA (kcal/mol)\\
    \midrule 
    DB12703 &Omipalisib &-13.06 \\
DB11896 &Gedatolisib &-12.36 \\
DB13109 &PKI-179 &-12.34 \\
DB11836 &Sapanisertib &-12.20 \\
DB12774 &AZD-8055 &-11.74 \\
DB11925 &Vistusertib &-11.50 \\
DB13072 &GDC-0349 &-11.24 \\
DB06233 &Ridaforolimus &-11.21 \\
DB12570 &CC-223 &-11.01 \\
    \bottomrule
    \end{tabular} 
    }
\end{table}

\begin{figure}[H]
	\centering
	\includegraphics[width = 1\textwidth]{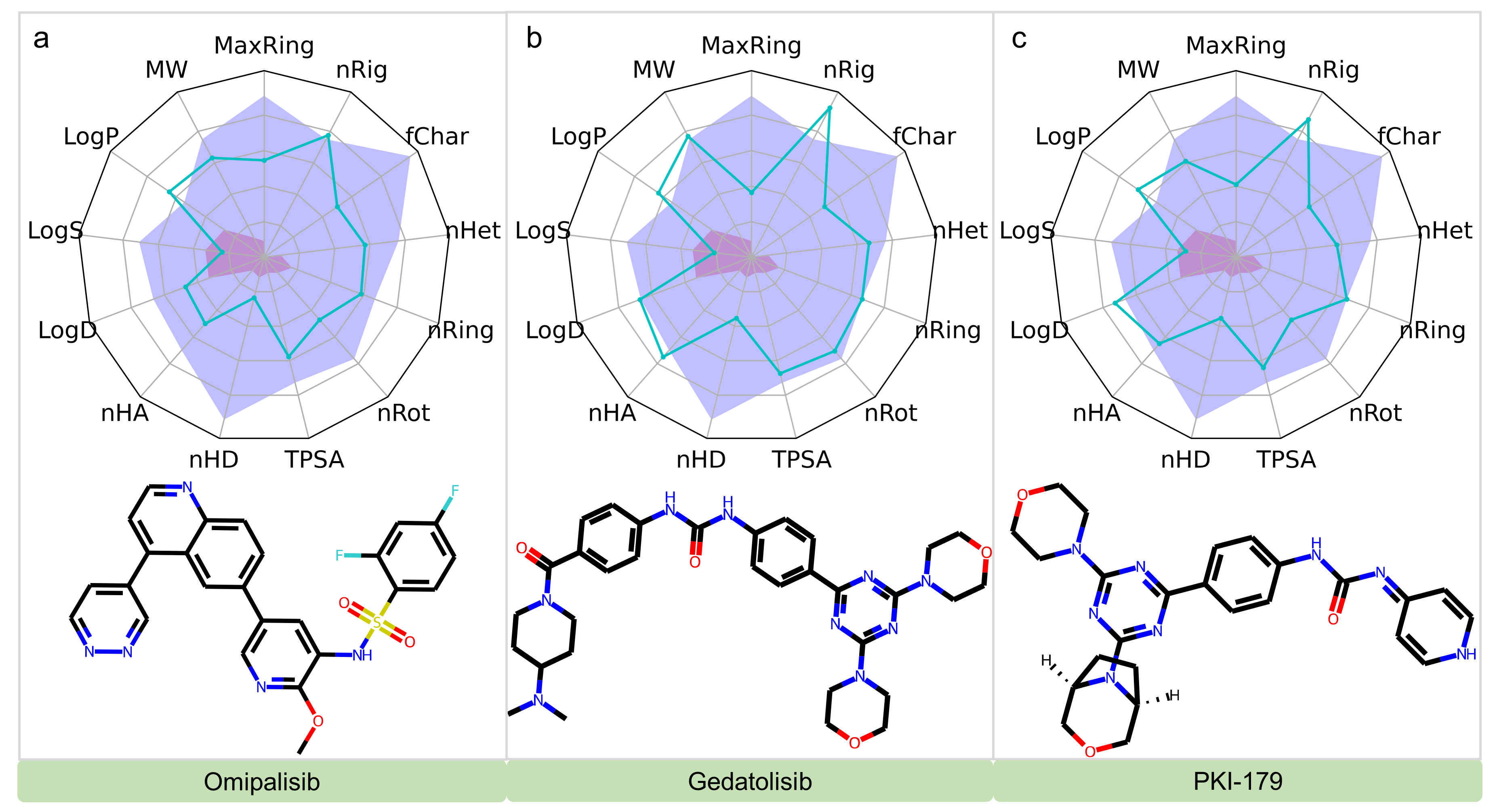}
	\caption{Evaluations of ADMET Properties for Omipalisib, Gedatolisib, and PKI-179: This figure illustrates the ADMET profiles of Omipalisib, Gedatolisib, and PKI-179, with the blue curves representing the values of 13 specified ADMET properties. The yellow and red zones demarcate the upper and lower limits, respectively, of the optimal ranges for these properties. The data shown are predictive results obtained from the ADMETlab 2.0 website (https://admetmesh.scbdd.com/). The evaluated ADMET properties include Molecular Weight (MW), the logarithm of the octanol/water partition coefficient (logP), the logarithm of the aqueous solubility (logS), the logP at physiological pH 7.4 (logD), Number of hydrogen bond acceptors (nHA), Number of hydrogen bond donors (nHD), Topological polar surface area (TPSA), Number of rotatable bonds (nRot), Number of rings (nRing), Number of atoms in the biggest ring (MaxRing), Number of heteroatoms (nHet), Formal charge (fChar), and Number of rigid bonds (nRig).}
	\label{fig:mtor_admet}
\end{figure}

\paragraph{Approved drugs with predicted efficacy on mGluR5}

In our analysis of DrugBank, we found that fewer approved drugs exhibited a high binding affinity for mGluR5 as compared to mTOR. Specifically, only two drugs, Doravirine and Desogestrel, displayed a binding affinity better than -10 kcal/mol, as detailed in Table \ref{tab: mGluR5_drug}.
\begin{table}[H] 
    \centering
    \caption{Summary of the FDA-approved drugs that are potential potent inhibitors of mGluR5 with BA < -9.54 kcal/mol.}
    \label{tab: mGluR5_drug}
    {
    \begin{tabular}{ ccc } 
    \toprule
    DrugID	&Name&	Predicted BA (kcal/mol)\\
    \midrule 
    DB12301 &Doravirine &-10.52 \\
    DB00304 &Desogestrel &-10.05 \\
    \midrule
    DB12612 &Ozanimod &-9.90 \\
    DB01595 &Nitrazepam &-9.87 \\
    DB06636 &Isavuconazonium &-9.85 \\
    DB09291 &Rolapitant &-9.78 \\
    DB08439 &Parecoxib &-9.75 \\
    DB00294 &Etonogestrel &-9.73 \\
    DB00475 &Chlordiazepoxide &-9.72 \\
    DB11633 &Isavuconazole &-9.68 \\
    DB00404 &Alprazolam &-9.65 \\
    DB00904 &Ondansetron &-9.63 \\
    DB15685 &Selpercatinib &-9.62 \\
    DB11374 &Amprolium &-9.57 \\
    
    \bottomrule
    \end{tabular} 
    }
\end{table}

Doravirine serves as an antiviral agent specifically designed to combat HIV infections. It functions as a non-nucleoside reverse transcriptase inhibitor (NNRTI), targeting the reverse transcriptase enzyme of HIV-1. This enzyme is central to HIV replication, as it synthesizes complementary DNA (cDNA) from the virus's RNA. Importantly, Doravirine's mode of action non-competitively inhibits the HIV-1 reverse transcriptase without affecting human cellular DNA polymerases $\alpha$, $\beta$ , and mitochondrial DNA polymerase $\gamma$ \cite{lai2014vitro}. Desogestrel is a synthetic progestin used primarily for contraception, either alone or in combination with an estrogen such as ethinyl estradiol. Its contraceptive efficacy arises mainly from the inhibition of ovulation\cite{scala2013drug}. Ozanimod, commercially known as Zeposia, is a medication used for the treatment of relapsing forms of multiple sclerosis\cite{swallow2020comparative} and ulcerative colitis\cite{paik2022ozanimod} in adults. Its mechanism of action is centered on modulating the sphingosine-1-phosphate (S1P) receptor. This modulation curtails the proliferation of lymphocytes, which are pivotal in inciting inflammation within the cerebral and gastrointestinal regions.

\begin{figure}[H]
	\centering
	\includegraphics[width = 1\textwidth]{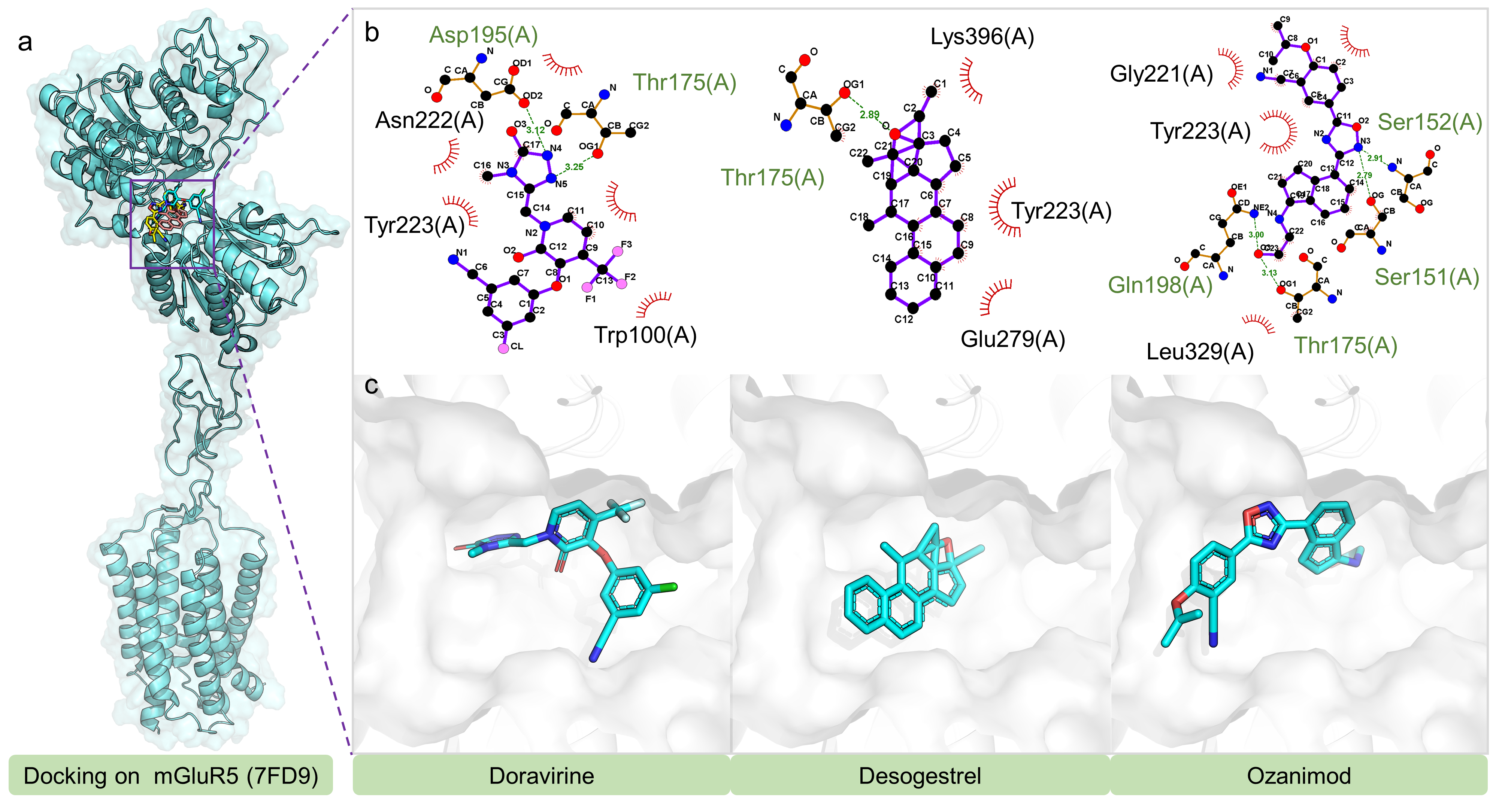}
	\caption{The docking structures and interactions of Doravirine, Desogestrel, and Ozanimod with mGluR5.}
	\label{fig:mglu5_dock}
\end{figure}

We performed a detailed exploration of the binding mechanisms of these three drugs with mGluR5, utilizing molecular docking techniques. Figure \ref{fig:mglu5_dock} illustrates the binding poses and intermolecular interactions between Doravirine, Desogestrel, and Ozanimod with the mGluR5 active site. A salient feature observed across all three drug-receptor complexes is the hydrogen bonding interaction with the Thr175 residue. Specifically, Doravirine forms this bond utilizing its oxygen atom, whereas Desogestrel and Ozanimod achieve this interaction through their respective nitrogen atoms. This consistent interaction with Thr175 underscores its potential as a critical residue in modulating drug-target affinity within the mGluR5 receptor. Further interaction analysis revealed that Doravirine establishes an additional hydrogen bond with Asp195. In contrast, Ozanimod predominantly engages via hydrogen bond interactions, interacting notably with Ser151, Ser152, and Gln198. Desogestrel, in its interaction profile, primarily engages in hydrophobic interactions, emphasizing its distinct binding dynamics within the mGluR5 receptor's active site. 

\paragraph{Investigational drugs with predicted efficacy on mGluR5}
In our subsequent analysis, we explored further investigational drugs projected to influence mGluR5 activity. It was observed that a greater number of these investigational drugs exhibited a higher affinity towards mGluR5 compared to approved drugs. Table \ref{tab: mGluR5_investigational} restrictively lists compounds boasting a binding affinity stronger than -10kcal/mol.
\begin{table}[H] 
    \centering
    \caption{Summary of investigational drugs that have the potential to inhibit mGluR5.}
    \label{tab: mGluR5_investigational}
    {
    \begin{tabular}{ ccc } 
    \toprule
    DrugID	&Name&	Predicted BA (kcal/mol)\\
    \midrule 
    DB13004 &Mavoglurant &-10.88\\
    DB12733 &Dipraglurant &-10.82\\
    DB11649 &Lersivirine &-10.40\\
    DB12999 &MK-6186 &-10.17\\
    DB15406 &GLPG-0974 &-10.11\\
    DB04885 &Cilansetron &-10.11\\
    DB14929 &Elsulfavirine &-10.10\\
    DB13035 &AG-24322 &-10.02\\
    DB12931 &Fenobam &-10.01\\
    \bottomrule
    \end{tabular} 
    }
\end{table}
Among these, Mavoglurant and Dipraglurant stand out as mGluR5 antagonists. Mavoglurant's early investigations were centered around its potential treatment for Fragile X syndrome\cite{gomez2014development}, and subsequently, for levodopa-induced dyskinesia (LID) associated with Parkinson's disease (PD)\cite{negida2021mavoglurant}. Unfortunately, its progression to clinical use has been halted due to its inability to achieve primary endpoints in advanced clinical trials. On the other hand, Dipraglurant has been specifically formulated to mitigate LID prevalent in PD patients\cite{tison2016phase}. Lersivirine (or UK-453061) is a distinct compound from the pyrazole class, emerging as a part of the next-generation NNRTIs. Spearheaded by ViiV Healthcare, its primary purpose was to challenge HIV-1 infection\cite{platten2013lersivirine}. 

\begin{figure}[H]
	\centering
	\includegraphics[width = 1\textwidth]{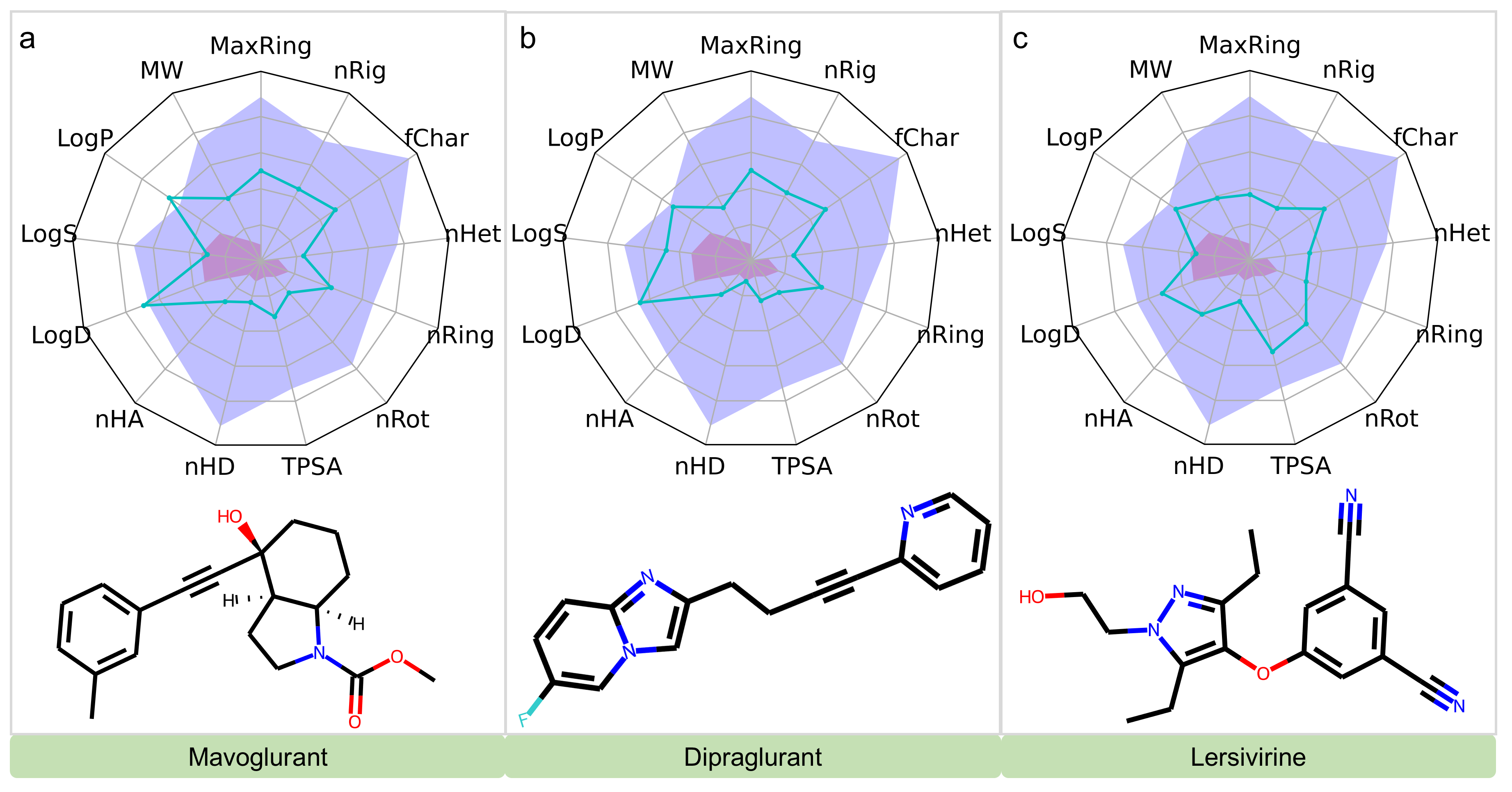}
	\caption{Evaluations of ADMET Properties for Mavoglurant, Dipraglurant, and Lersivirine: This figure showcases the ADMET profiles for Mavoglurant, Dipraglurant, and Lersivirine. The blue curves in the graph indicate the values for 13 specific ADMET properties of these compounds. The yellow and red zones in the graph are designated to highlight the upper and lower limits of the optimal ranges for each of these ADMET properties, respectively.}
	\label{fig:mglu5_admet}
\end{figure}

\paragraph{Approved drugs with predicted efficacy on NMDAR}
In our analysis of DrugBank, we identified only 7 approved drugs that exhibit a binding affinity greater than -9.54 kcal/mol to NMDAR. (Table \ref{tab: NMDAR_drug}) 
\begin{table}[H] 
    \centering
    \caption{Summary of the FDA-approved drugs that are potential potent inhibitors of NMDAR with BA < -9.54 kcal/mol.}
    \label{tab: NMDAR_drug}
    {
    \begin{tabular}{ ccc } 
    \toprule
    DrugID	&Name&	Predicted BA (kcal/mol)\\
    \midrule 
    DB00157 &NADH &-10.10 \\
DB03147 &FAD &-10.09 \\
\midrule 
DB00705 &Delavirdine &-9.75 \\
DB00131 &Adenosine phosphate &-9.69 \\
DB00118 &Ademetionine &-9.61 \\
DB00364 &Sucralfate &-9.56 \\
DB08874 &Fidaxomicin &-9.56 \\
    \bottomrule
    \end{tabular} 
    }
\end{table}
Among these, Nicotinamide Adenine Dinucleotide (NADH) and Flavin Adenine Dinucleotide (FAD) stand out as crucial coenzymes that participate in various biochemical reactions within the body. There is some evidence to suggest that NADH could be beneficial in treating conditions such as Parkinson's disease and chronic fatigue syndrome\cite{birkmayer1993nicotinamide}. Delavirdine, marketed under the brand name Rescriptor, is classified as a NNRTI and was developed to treat HIV-1\cite{scott2000delavirdine}. It exerts its therapeutic effects by binding directly to the reverse transcriptase enzyme, thereby inhibiting both RNA-dependent and DNA-dependent DNA polymerase activities. Adenosine phosphate is an adenine nucleotide with a single phosphate group. Initially introduced as a vasodilator and anti-inflammatory agent, it faced withdrawal by the FDA due to concerns regarding its safety and efficacy for these indications. Nevertheless, it has found applications in nutritional supplementation and in addressing dietary insufficiencies or imbalances\cite{rao2023effects}. Ademetionine, commonly referred to as S-adenosylmethionine (SAMe), is a naturally occurring molecule in the body. It serves as a crucial physiological methyl donor, participating in enzymatic transmethylation processes essential to all living organisms. In the U.S., SAMe has been marketed as a dietary supplement, touted for its benefits in supporting mood and emotional well-being\cite{bressa1994s}.

\begin{figure}[H]
	\centering
	\includegraphics[width = 1\textwidth]{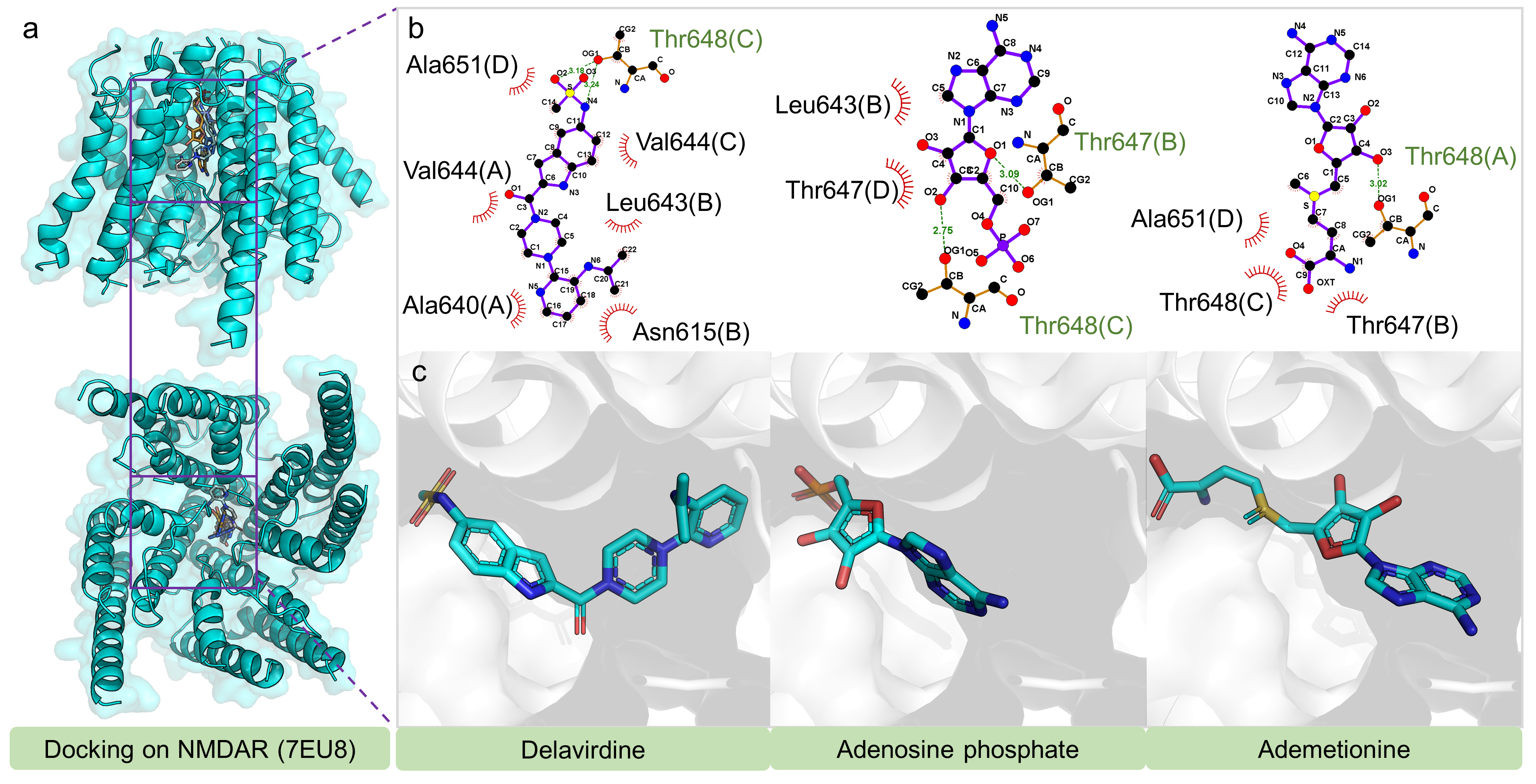}
	\caption{The docking structures and interactions of Delavirdine, Adenosine phosphate, and Ademetionine with NMDAR.}
	\label{fig:NMDAR_dock}
\end{figure}

We conducted docking studies of Delavirdine, Adenosine phosphate, and Ademetionine with the active site of NMDAR to elucidate their interaction patterns (Figure \ref{fig:NMDAR_dock}). Predominantly, these drugs employ hydrophobic interactions for their binding to NMDAR. Notably, all three drugs establish hydrogen-bonding interactions with Thr648. Furthermore, Adenosine phosphate uniquely forms a hydrogen bond with the side chain hydroxyl group of the neighboring residue, Thr647. We did not observe any other significant interaction types in this analysis.

\paragraph{Investigational drugs with predicted efficacy on NMDAR}
In our pursuit to identify potential modulators of NMDAR, we explored further the realm of investigational compounds. Notably, a larger proportion of these compounds exhibited higher affinity for NMDAR in comparison to approved drugs. For brevity, only the top 10 of these compounds are presented in Table \ref{tab: NMDAR_investigational}.
\begin{table}[H] 
    \centering
    \caption{Summary of top-10 investigational drugs that have the potential to inhibit NMDAR.}
    \label{tab: NMDAR_investigational}
    {
    \begin{tabular}{ ccc } 
    \toprule
    DrugID	&Name&	Predicted BA (kcal/mol)\\
    \midrule 
    DB06741 &Gavestinel &-11.70 \\
DB12365 &Perzinfotel &-10.84 \\
DB12749 &Butylphthalide &-10.47 \\
DB05553 &Regrelor &-10.23 \\
DB12140 &Dilmapimod &-10.14 \\
DB03708 &Adenosine 5'-phosphosulfate &-10.12 \\
DB13019 &Henatinib &-9.92 \\
DB06334 &Tucidinostat &-9.84 \\
DB12012 &PF-04457845 &-9.83 \\
DB05973 &Inosine 5'-sulfate &-9.79 \\
    \bottomrule
    \end{tabular} 
    }
\end{table}
Among them, Gavestinel and Perzinfotel emerged as notable NMDAR antagonists. Gavestinel was originally earmarked for the therapeutic management of acute intracerebral hemorrhage. Mechanistically, it operates as a non-competitive antagonist, targeting the glycine binding site on the NMDA receptor. Gavestinel's specificity is underscored by its remarkable selectivity, manifesting over a 1000-fold preference for the NMDA receptor over other receptor binding sites like AMPA and kainate\cite{di1997substituted}. Beyond its in vitro affinity, Gavestinel also proved to be orally bioavailable with in vivo activity\cite{di1997substituted}. Perzinfotel, on the other hand, has been the subject of clinical investigations geared towards stroke\cite{kinney1998design}. Butylphthalide is a chemical constituent isolated from celery oil. Preliminary research, particularly in preclinical models, has hinted at its multifaceted therapeutic potential in hypertension management and neuroprotection\cite{chen2019application}. The neuroprotective potential of butylphthalide garnered considerable attention in clinical circles, especially its therapeutic potential against acute ischemic stroke\cite{wang2023efficacy}.

\begin{figure}[H]
	\centering
	\includegraphics[width = 1\textwidth]{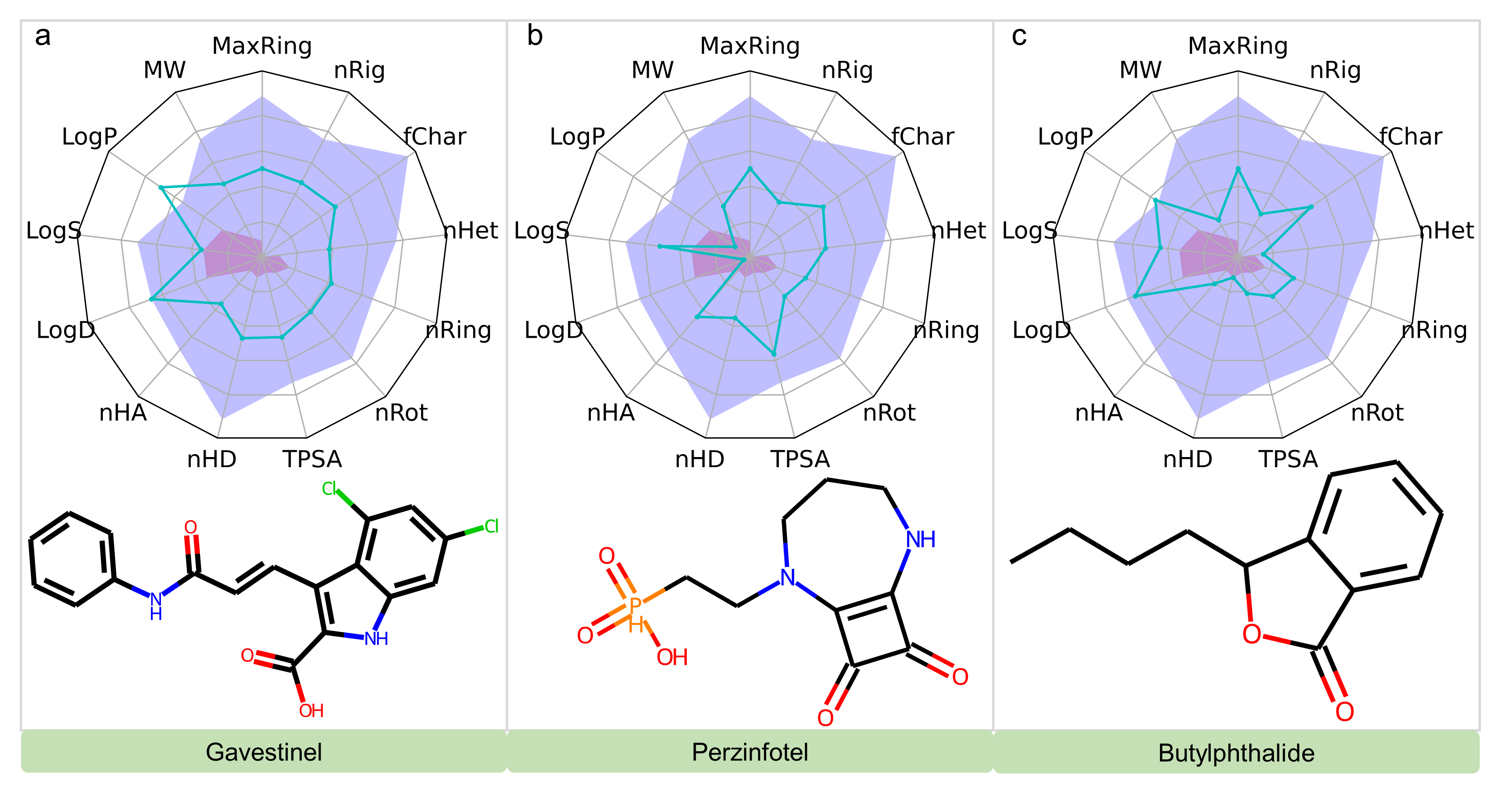}
	\caption{Evaluations of ADMET Properties for Gavestinel, Perzinfotel, and Butylphthalide.}
	\label{fig:NMDAR_admet}
\end{figure}

\subsubsection{ADMET analysis}
Understanding that a molecule's pharmacokinetic and safety profiles are pivotal in its clinical trial success, we focused on the ADMET (absorption, distribution, metabolism, excretion, and toxicity) properties of our identified investigational drugs. A comprehensive understanding of these properties is crucial, as drug candidates can fail in advanced stages if they do not meet ADMET criteria, which govern their behavior within the human body\cite{xiong2021admetlab}. Utilizing the ADMETlab 2.0 tool, we assessed 13 distinct ADMET attributes for each drug which show potential efficacy on the three targets\cite{xiong2021admetlab}. 

For mTOR inhibitors, our analysis revealed that Omipalisib has a logP value slightly exceeding the optimal range, which can potentially affect its solubility and permeability (Figure \ref{fig:mtor_admet}). However, its other ADMET properties fell within acceptable limits, suggesting a favorable pharmacokinetic profile overall. On the other hand, Gedatolisib presented some challenges, with a relatively large molecular weight and a higher number of rings and hydrogen bond donors. These structural features contribute to its high logP value, resulting in poor water solubility. Similarly, while PI3K, developed from Gedatolisib, has a smaller molecular weight, it inherits the challenge of a high logP value and limited water solubility.

In the case of mGluR5 inhibitors (Figure \ref{fig:mglu5_admet}), we found that Mavoglurant's logP and logD values surpassed recommended thresholds, indicating potential issues with water solubility. However, its other ADMET properties were within acceptable ranges. Dipraglurant and Lersivirine, on the other hand, showed well-balanced ADMET profiles, suggesting favorable drug-like characteristics for these compounds.

For potential NMDAR therapeutic agents, Gavestinel, Perzinfotel, and Butylphthalide, were also systematically predicted (Figure \ref{fig:NMDAR_admet}). The results revealed that all three compounds exhibited favorable ADMET properties. Gavestinel, with its moderate molecular weight, showed a slightly elevated logP value attributed to its hydrophobic groups, leading to marginal water solubility limitations. Despite this, Gavestinel retained its oral bioavailability. Contrarily, Perzinfotel exhibited enhanced water solubility due to the presence of a phosphate group within its molecular structure, which concomitantly resulted in reduced logP and logD values. Lastly, Butylphthalide presented with a logP value on the verge of the upper limit, yet all other properties fell within the scientifically recommended and acceptable range, demonstrating its optimal characteristics.

In conclusion, this comprehensive ADMET analysis offers crucial insights into the pharmacokinetic and safety profiles of these investigational drugs. These findings are instrumental in guiding their further development and optimization for potential use in treating substance addiction.

\section{Methods}

\subsection{DEG analysis and PPI network}

The gene expression datasets for our DEG analysis were obtained from the GEO database. For the statistical analysis, we utilized the Limma\cite{ritchie2015limma} package to identify DEGs. For opioid addiction, the dataset GSE87823 was used. To minimize potential biases from age and gender, we included only male samples aged between 19 and 25. The gene expression data underwent quantile normalization. The set criteria for significance were a |Log2FoldChange| of 2 and a p-value threshold of 0.01. In the analysis of cocaine addiction, we employed the dataset GSE54839\cite{bannon2014molecular}. All samples were incorporated since control subjects were appropriately matched with chronic cocaine abusers. It is noteworthy that quantile normalization had already been applied to this dataset by its provider. The thresholds applied were a |Log2FoldChange| of 0.2630344 and a p-value of 0.05. In instances where genes appeared multiple times, we retained only the probe with the highest average expression across samples. Probes associated with multiple genes were excluded. 
The PPI network among the DEGs was sourced from the STRING database\cite{szklarczyk2023string}. To facilitate a multi-resolution analysis for each gene set, we employed four distinct interaction confidence thresholds: 0.15, 0.4, 0.7, and 0.9. 

\subsection{Multiscale topological differentiation of network}

The analysis of PPI networks presents significant challenges due to their inherent complexity and the dynamic nature of protein interactions.  Their analysis is complicated by factors such as the high-dimensional space they occupy and the variability in interaction strengths. Traditional methods often struggle to capture the nuanced relationships and structural variances within these networks, leading to incomplete or oversimplified interpretations. To overcome these challenges, we propose a multiscale and multi-resolution topological differentiation approach, which leverage the power of PST (or persistent Laplacians) and persistent homology (PH) to analyze the PPI network at various scales, resolutions, and dimensions. By closely examining the difference in topological invariants and geometric features, we can infer the significance of each removed node in a network. Significant alterations in these features typically indicate a high importance of the deleted gene in the network's overall functionality.

\subsubsection{Persistent spectral theory}
The PPI network is conceptualized as a graph where nodes correspond to proteins, and edges denote the interactions between pairs of proteins. Extending beyond the graph structure, we employ the concept of a simplicial complex to allow high-order interactions in the network,  
encapsulating  a more comprehensive range of shapes and facilitating the inclusion of high-dimensional topological relationships. Persistent spectral theory can capture topological and geometric changes over a continuous range of spatial resolutions, thus revealing the multiscale structure of the data. This is performed by constructing a filtration, which is essentially a nested sequence of simplicial complexes, indexed by a parameter $t$ that typically represents a scale or threshold level. In mathematical terms, a filtration of an oriented simplicial complex $K$ is a collection of sub-complexes $(K_t)_{t\in \mathbb{R}^+}$ satisfying:
\begin{equation}
    \emptyset = K_0 \subseteq K_1 \subseteq K_2 \subseteq \cdots \subseteq K_m = K.
\end{equation}
Here, $K$ is the largest simplicial complex achievable from the filtration, with each $K_t$ being a full simplicial complex at the filtration level $t$.

Accompanying this filtration is a corresponding sequence of chain complexes and boundary operators at each scale $t$:
\begin{equation}
    \cdots C_{q+1}^t 
            \xrightleftharpoons[\partial_{q+1}^{t^\ast}]{\partial_{q+1}^t} C_q^t 
            \xrightleftharpoons[\partial_q^{t^\ast}]{\partial_q^t}  \cdots 
            \xrightleftharpoons[\partial_3^{t^\ast}]{\partial_3^t}  C_2^t  \xrightleftharpoons[\partial_2^{t^\ast}]{\partial_2^t}  C_1^t  \xrightleftharpoons[\partial_1^{t^\ast}]{\partial_1^t}  C_0^t  \xrightleftharpoons[\partial_0^{t^\ast}]{\partial_0^t}  \varnothing, 
\end{equation}
where $C_q^t$ denotes the chain group for the sub-complex $K_t$ and $\partial_q^t$ is the $q$-th boundary operator mapping from $C_q^t$ to $C_q^{t-1}$. For each $K_t$, every $q$-simplex is oriented, and the boundary operator $\partial_q^t$ is applied as follows:
\[
    \partial_q^t(\sigma_q) = \sum_{i}^q(-1)^i[v_0, \cdots, \hat{v_i} ,\cdots,v_q], \sigma_q \in K_t.
    \]
Here, $\sigma_q = [v_0, \cdots, v_q]$ is an oriented $q$-simplex within $K_t$, and $[v_0, \cdots, \hat{v_i} ,\cdots,v_q]$ represents the oriented $(q-1)$-simplex obtained by omitting the $i$-th vertex $v_i$ from $\sigma_q$.

Consider $\mathbb{C}_{q}^{t+p}$ to be the subset of $C_q^{t+p}$ consisting of chains whose boundaries are in $C_{q-1}^t$:
\begin{equation}
    \mathbb{C}_q^{t+p} \coloneqq \{ \alpha \in C_q^{t+p} \ | \ \partial_q^{t+p}(\alpha) \in C_{q-1}^{t}\}.
\end{equation}
The $\eth_q^{t+p}$ is defined by:
\begin{equation}\label{equ: new boundary map}
    \eth_q^{t+p} : \mathbb{C}_q^{t+p} \to  C_{q-1}^{t}.
\end{equation}
The $p$-persistent $q$-combinatorial Laplacian operator $\Delta_q^{t+p}$ along the filtration is given by:
\begin{equation}
    \Delta_q^{t+p} = \eth_{q+1}^{t+p} \left(\eth_{q+1}^{t+p}\right)^\ast + \partial_q^{t^\ast} \partial_q^t.
\end{equation}
The matrix representations of the boundary operators $\eth_{q+1}^{t+p}$ and $\partial_q^t$ are $\mathcal{B}_{q+1}^{t+p}$ and $\mathcal{B}_{q}^t$, respectively. The ${p}$-persistent ${q}$-combinatorial Laplacian matrix $\mathcal{L}_q^{t+p}$ is defined as:
\begin{equation}\label{equ:PLC}
    \mathcal{L}_q^{t+p} = \mathcal{B}_{q+1}^{t+p} (\mathcal{B}_{q+1}^{t+p})^T + (\mathcal{B}_{q}^t)^T \mathcal{B}_{q}^t.
\end{equation}
This matrix is symmetric and positive semi-definite, ensuring that all eigenvalues are real and non-negative. The $p$-persistent $q$th Betti numbers, representing the number of $q$-cycles persisting in $K_t$ after $p$ filtration, correspond to the nullity of $\mathcal{L}_q^{t+p}$:
\begin{equation}\label{equ:persistent betti}
    \beta_q^{t+p} = \text{dim}(\mathcal{L}_{q}^{t+p}) - \text{rank}(\mathcal{L}_{q}^{t+p}) = \text{nullity}(\mathcal{L}_{q}^{t+p}) = \boldsymbol{\#} ~{\rm of ~zero ~eigenvalues ~of ~}   \mathcal{L}_{q}^{t+p}.
\end{equation}
PST provides geometric insights from the spectra of this persistent combinatorial Laplacian, extending beyond mere topological persistence. Persistent Betti numbers offer information on topological constancy, while geometric transformations are discernible through the non-harmonic portions of the spectrum.

\subsubsection{Persistent homology}

Complementing PST, PH offers a different approach for multiscale analysis through filtration techniques \cite{zomorodian2004computing}. In this framework, for a given simplicial complex $K$, we define the $q$-cycle group $\mathcal{Z}_q$ and the $q$-boundary group $\mathcal{B}_q$ as the kernel and image of boundary operators $\partial_q$ and $\partial_{q+1}$, respectively. The equations

\begin{eqnarray}
    \mathcal{Z}_q&=& \mbox{Ker}\partial_q=\{c \in {C}_q\mid\partial_qc = 0\},\\ \mathcal{B}_{q}&=& \mbox{Im}\partial_{q+1}=\{c\in {C}_q | \exists d\in    {C}_{q+1}: c= \partial_{q+1}d
    \},\end{eqnarray}
describe these groups, where ${C}_q$ represents the set of $q$-chains. Since $\partial_q \circ \partial_{q+1} = 0$, it follows that $\mathcal{B}_q \subseteq \mathcal{Z}_q \subseteq {C}_q$. Consequently, we define the $q$-homology group $\mathcal{H}_q$ as the quotient group:
\begin{equation}
    \mathcal{H}_q = \mathcal{Z}_q / \mathcal{B}_q.
    \end{equation}
The $q$th Betti number $\beta_q$ is then the rank of $\mathcal{H}_q$. Furthermore, the persistence of these topological features is quantified by the $q$th persistence Betti number $\beta^{i,j}_q$, which is the rank of the homology groups of $K_i$ that persist to $K_j$, formulated as

\begin{equation}
\beta^{i,j}_q = \text{rank}(\mathcal{Z}_q(K_i) / (\mathcal{B}_q(K_j) \cap \mathcal{Z}_q(K_i))).
\end{equation}

Through filtration, PH provides a detailed perspective on the persistence of topological invariants, though it primarily focuses on the harmonic spectral aspects of PST.

\subsubsection{Key gene identification via network topological differentiation}
We can make use of  PST and/or PH for topological differentiation analysis to assess the significance of individual genes within a PPI network. Using the STRING database, we quantify the interaction strength between protein pairs with a combined score. This score serves as a basis for formulating an abstract distance between proteins, which facilitates the construction of a Rips complex. To conduct a multi-resolution and multiscale analysis, we utilize four PPI network using thresholds of 0.15, 0.4, 0.7, and 0.9. The network can be denoted as $G=(V,E)$. Within this framework, we use a concept known as `topological perturbation analysis', proposed by Chen et al.\cite{chen2023path}, at the $m$-th vertex $v_m$ by considering the subgraph $G_m$ that results from eliminating $v_m$ and all edges linked to $v_m$. For a specified threshold $T$, we define the distance between two proteins $v_i$ and $v_j$ in $G$ as:
\begin{equation}\label{distance matrix}
    D_{ij} = 
    \begin{cases}
        1-s_{ij},   & \mbox{if $s_{ij} > T$.} \\
        \infty ,                         & \text{otherwise.} 
    \end{cases}
\end{equation}
where $s_{ij}$ denote the combined score of interaction between proteins $v_i$ and $v_j$ in STRING database. In this study, PST is utilized to derive both topological and geometric characteristics of the network in its original and perturbed states. For the analysis, a Rips complex-based filtration up to two dimensions is established. In each network case, ten filtration parameters are evenly chosen from the interval $0$ to $(1-T)$. From each Laplacian spectrum, we compute the harmonic spectra count and the five statistical descriptors (minimum, maximum, mean, standard deviation, and sum) of the non-harmonic spectra. For every network $G$, we encapsulate its attributes into a vector $f_G$, defined as $f_G=\Theta (G)$. Correspondingly, after perturbation at vertex $v_m$, we obtain the feature vector $f_{G_m}$ for each $G_m$, defined as $f_{G_m}=\Theta (G_m)$, where $\Theta$ encapsulates PST analysis and network vectorization. We quantify the significance of the node $v_m$ by computing the Euclidean distance between feature vectors $f_G$ and $f_{G_m}$:
\begin{equation}\label{significance}
    S_{m}^G = distance(f_G,f_{G_m}).
\end{equation}
This metric reflects the impact of gene $v_m$ within the PPI network $G$ by indicating how its removal alters the network's structure in terms of topology and geometry.

\subsection{Machine learning-based drug repurposing}

\subsubsection{Data preparation}
For our machine learning models, we sourced inhibitor datasets pertaining to mTOR, mGluR5 and NMDAR from the ChEMBL database\cite{mendez2019chembl}. These datasets comprise SMILES strings of molecular compounds, each paired with a bioactivity label. The initial labels assigned to these data points were either $\text {IC}_{50}$ or $K_i$ values. To adapt these experimental labels into binding affinities (BAs) suitable for our models, we employed the conversion formula:  BA $= 1.3633 \times \log_{10}{K_i} ($kcal/mol). $\text {IC}_{50}$ labels were subsequently estimated to $K_i$ values based on the relationship $K_i = \text {IC}_{50}/2$, in alignment with recommendations by Kalliokoski\cite{kalliokoski2013comparability}. For instances where a single molecule had multiple bioactivity value labels, we computed the average of these labels. In addition, we retrieved small molecule drugs, categorized under either approved or investigational status, from the DrugBank database (version 5.1.10)\cite{wishart2018drugbank}. To ensure consistency, their SMILES strings were canonicalized by the RDKit toolkit. 

\subsubsection{Molecular fingerprints}
In our molecular analysis, we employed three fingerprinting methodologies to delineate molecular structures. Two of them harness advanced NLP techniques: one utilizing a bidirectional transformer-based model\cite{chen2021extracting} and the other deploying a sequence-to-sequence autoencoder framework\cite{winter2019learning}. These NLP-driven methodologies leverage pretrained models to transduce canonical SMILES notations into 512-dimensional latent vectors. Complementing these, our study also incorporated a classical topological approach characterized by the 2D ECFP\cite{rogers2010extended}, synthesized via the comprehensive RDKit computational toolkit.

\paragraph{Bidirectional transformer}
Chen and colleagues innovatively crafted a self-supervised learning (SSL) framework for the purpose of pretraining deep neural networks using millions of unlabeled molecular structures\cite{chen2021extracting}. This methodology yields latent vectors derived from input SMILES, encapsulating essential molecular structural details, thereby serving as robust fingerprints for subsequent machine learning applications. The core of their SSL platform employs the bidirectional encoder transformer (BET) model, which leverages attention mechanisms for enhanced accuracy. During the pretraining phase on the SSL platform, the process involved the formation of data pairs comprising authentic SMILES strings and their masked counterparts, with a specific fraction of symbols intentionally obscured. To create these pairs, 15\% of the symbols in all SMILES strings were masked. In this masking process, 80\% of the symbols were fully obscured, 10\% were left unaltered, and the remaining 10\% were randomly modified. The employment of the attention mechanism in the BET model ensures that the significance of each symbol within the SMILES string is fully captured. In their study, Chen et al. utilized SMILES strings extracted from one or a combination of the ChEMBL\cite{mendez2019chembl}, PubChem\cite{kim2016pubchem}, and ZINC\cite{irwin2005zinc} databases for training their SSL-based BET model. In the context of the current study, we opted to utilize the fingerprints generated directly from the pretrained model on ChEMBL database, without any additional fine-tuning (termed as TF-FPs). 

\paragraph{Sequence-to-sequence auto-encoder}
A novel unsupervised learning approach, utilizing a sequence-to-sequence autoencoder, has been developed to interpret molecular information encapsulated within the SMILES representation\cite{winter2019learning}. By translating one molecular format into another, the model efficiently compresses detailed chemical structures into a latent space positioned between the encoder and decoder components. During this transformation, intermediary vectors embed significant physicochemical details, enabling the pretrained model to retrieve molecular descriptors from input SMILES strings without additional training. The encoder employs a combination of convolutional neural network (CNN) and recurrent neural network (RNN) structures, directing their outputs to form intermediary vector representations. On the other hand, the decoder, which is mainly based on RNN architectures, interprets these vectors to produce the desired output. To enhance the richness of the latent vectors, an auxiliary classification module is incorporated, forecasting specific molecular properties. The loss function, instrumental during the autoencoder model's training, is a fusion of cross-entropies associated with the decoder's probabilistic character outputs and mean squared errors pertinent to molecular property predictions. This comprehensive model has been rigorously trained on expansive datasets sourced from the ZINC and PubChem databases.

\paragraph{Extended-connectivity fingerprints}
Extended-connectivity fingerprints (ECFPs) are a unique class of topological fingerprints used for molecular characterization, primarily developed for structure-activity modeling\cite{rogers2010extended}. The ECFP approach assigns specific identifiers to each atom in a molecule based on their properties and surrounding atoms. Through iterative hashing techniques, these identifiers are updated to encompass the molecule's neighborhood information, and redundant features are removed. Ultimately, these identifiers are consolidated into a bit array, capturing the molecule's distinctive features. We employed the RDKit library to produce ECFPs. This library constructs circular fingerprints using the Morgan algorithm, which necessitates a radius parameter dictating the algorithm's iteration count. In our study, the ECFP radius was set to 2, with the fingerprint length designated at 2048. 

\subsubsection{Machine learning models}

We adopted the GBDT algorithm for constructing our machine learning models\cite{friedman2001greedy, ke2017lightgbm}. GBDT can be utilized for both regression and classification tasks. The algorithm operates by initially utilizing a weak learner, predominantly a decision tree, to offer primary predictions, subsequently quantifying the residual errors between these initial predictions and the actual outcomes. A subsequent decision tree is tailored to model these residuals, with the explicit objective of correcting inaccuracies introduced by the preceding trees in the sequence. This iterative methodology continues, with each iteration aiming to refine the residuals from the preceding step, until the total tree count attains a predetermined number or the cumulative residuals satisfy a specified threshold. The final model is an aggregate of all the individual trees, which are used for inference on new datasets.

In our investigation, we utilized three molecular fingerprint modalities, namely TF-FP, AE-FP, and ECFP, to represent inhibitor molecules. Three distinct machine learning models were subsequently curated using the GBDT framework. To augment the predictive accuracy and robustness of the models in assessing BAs for mTOR, mGluR5 and NMDAR, a consensus-based approach was adopted. This consensus model combined the predictions by averaging the binding affinity predictions sourced from each of the three individual models.

\section{Conclusion}

The escalating crisis of drug addiction in the United States calls for effective therapeutic interventions. This study embarked on an comprehensive and rigorous strategy, moving from transcriptomic data analysis to drug discovery, with the goal of identifying potential repurposed drug candidates for opioid and cocaine addiction. The aim is to mitigate the severe societal and health detriments associated with these addictions.

Our sophisticated framework commenced with DEG (Differentially Expressed Gene) analysis on transcriptomic data. We propose a groundbreaking topological differentiation to identify key genes in a multiscale and highly reliable manner. The resulting targets are cross-validated through pathway analysis and literature review. These methodological advancements facilitated the identification of potential new targets, leading to the construction of a comprehensive machine learning repurposing study. We screen potential drugs using Transformer and autoencoder embeddings, along with traditional 2D fingerprints. This marks a pivotal transition from transcriptomic insights to actionable drug repurposing avenues. The resulting promising drug candidates are further screened for their absorption, distribution, metabolism, excretion, and toxicity (ADMET) properties. This critical evaluation of the pharmacokinetic and safety profiles of identified drug candidates underpins their therapeutic potential.

Through rigorous validation and analysis, three molecular targets, namely mTOR, mGluR5, and NMDAR, were identified as highly relevant to substance addiction. The machine learning models exhibited robust predictive capabilities in evaluating binding affinities, leading to the identification of promising drugs with satisfactory binding energies and favorable pharmacokinetic properties from DrugBank. However, the identified promising drugs need further in vivo validation to ascertain their safety and efficacy in mitigating substance addiction.

Our study signifies an advancement in employing computational methodologies, including advanced topological data analysis, to traverse the continuum from transcriptomic data to drug discovery. It lays down a robust framework for ensuing research aimed at unveiling potent therapeutic interventions, thus underscoring the vital role of interdisciplinary research in navigating the intricate landscape of drug addiction and advancing the drug discovery paradigm. The translational potential of our approach is vast, providing a scalable and adaptable framework that can be applied across a spectrum of diseases and transcriptomic datasets. The transcriptomic data utilized in this study originates from microarray analysis. However, the analytical methodology delineated herein can also be further applied to single-cell RNA sequencing (scRNA-Seq) data, thereby accounting for cellular heterogeneity within tissues. This acknowledgment of cellular heterogeneity is of paramount importance, particularly in the context of neurological diseases, where cell-specific variations significantly impact disease manifestation and progression. In conclusion, our study not only advances the field of addiction therapy but also sets a new benchmark for interdisciplinary research in drug discovery, demonstrating the potential to transform healthcare research paradigms.

\section{Data Availability}
The data and source code of this study are freely available at GitHub (https://github.com/Brian-hongyan/DEG-substance-addiction).

\section{Acknowledgements}
The work Wei was supported in part by NIH grants R01GM126189, R01AI164266, and R35GM148196, National Science Foundation grants DMS2052983 and IIS-1900473,   Michigan State University Research Foundation, and  Bristol-Myers Squibb  65109.

\section{Conflicts of Interests}
The authors declare no competing interests. 

\clearpage 


\newpage
\begin{appendices}\section{Literature validation}
    \subsection{Opioid addiction-related key genes}
    To gain deeper insights into the intricate relationship between our identified key genes and opioid addiction, we turned to literature validation. 
    
    One of the intriguing genes that surfaced was RHEB, whose association with opioid addiction appears to be mediated through the mTORC1 pathway. RHEB, a small GTPase, is instrumental in activating mTORC1\cite{ma2020persistent}. While RHEB's binding to mTOR occurs distally from the kinase's active site, it induces a pronounced global conformational shift \cite{yang2017mechanisms}. This allosteric modulation realigns active-site residues, enhancing catalytic actions. The significance of this interaction is underscored by the observed linkage between mTORC1 activation and opioid-induced phenomena such as tolerance and hyperalgesia. Diving deeper into the mechanisms, opioids have been postulated to induce alterations in protein translation within the nervous system\cite{xu2014opioid}. These changes are believed to set the stage for the emergence of tolerance and hyperalgesia. Xu et al. unearthed that after repeated morphine administrations, mTOR---a regulator of protein translation---becomes activated in rat spinal dorsal horn neurons\cite{xu2014opioid}. Notably, this mTOR activation is initiated through the $\mu$ opioid receptor and is channeled via the intracellular PI3K/Akt pathway. This discovery presents mTOR inhibitors as potential therapeutic agents in stymieing or diminishing opioid tolerance, particularly when addressing chronic pain. Furthermore, the overarching role of mTORC1 extends beyond opioids. Other substances, including cocaine \cite{bailey2012rapamycin, wu2011inhibition}, cannabinoids\cite{zubedat2017involvement}, and alcohol\cite{neasta2010role, neasta2014mtor, spanagel2009alcoholism}, have also been shown to activate mTORC1. This broad-spectrum influence underscores the pivotal role mTORC1 assumes in the realm of substance addiction. 
    
    ADCY9 is an integral protein implicated in the morphine addiction pathway, playing a pivotal role in the cellular signaling cascade. Its primary function is to catalyze the conversion of ATP into cyclic AMP (cAMP)\cite{hacker1998cloning}. One of the most noteworthy roles of cAMP is its ability to activate protein kinase A (PKA). When cAMP binds to the regulatory subunits of PKA, it triggers a conformational change, releasing the catalytic subunits. Upon release, these catalytic subunits are rendered active and embark on phosphorylating a myriad of protein targets inside the cell, initiating or modulating various cellular processes. A noteworthy aspect of PKA's function lies in its capability to modulate the dynamics of gamma-aminobutyric acid (GABA)\cite{couve2002cyclic, danielsson2016airway, kelm2008role, mcdonald1998adjacent}. PKA not only influences the secretion of GABA by neuronal cells\cite{danielsson2016airway, kelm2008role} but also plays a role in fine-tuning the functional responsiveness of GABA receptors\cite{couve2002cyclic, mcdonald1998adjacent}. GABA stands out as the principal inhibitory neurotransmitter within the central nervous system\cite{kumar2013therapeutic, sears2021influence}. It is instrumental in modulating neuronal excitability, thus maintaining a delicate balance between excitatory and inhibitory signals in the brain. A malfunction or dysregulation in ADCY9's activity could unleash a cascade of molecular events that may disturb this delicate balance. Such an aberration could potentially compromise GABAergic signaling, culminating in a state of hyperexcitability within the central nervous system. This heightened excitability is often accompanied by a surge in the release of dopamine, the neurotransmitter associated with pleasure and reward\cite{brodnik2018local, roberts2020gaba}.
    
    The adenomatous polyposis coli (APC) protein is intrinsically tied to opioid addiction through its interactions within the Wnt signaling pathway. Research has highlighted the Wnt pathway's involvement in withdrawal symptoms ensuing from opioid receptor activation, either due to exposure to morphine or as a result of chronic inflammation\cite{wu2020wnt}. Moreover, the manifestation of opioid-induced hyperalgesia, a paradoxical increase in pain sensitivity, has been tied to the activities of reactive astrocytes, which are intriguingly governed by Wnt5a signaling\cite{liu2023development}. APC plays a pivotal role in this milieu by acting as a negative regulator of the Wnt pathway. It is a key constituent of the destruction complex, a molecular assembly that also includes the proteins AXIN and Glycogen Synthase Kinase 3$\beta$ (GSK-3$\beta$)\cite{bugter2021mutations, liu2022wnt}. Within the destruction complex, GSK-3$\beta$ phosphorylates specific serine and threonine residues on $\beta$-catenin which marks $\beta$-catenin for ubiquitination and subsequent degradation.
    
    TBL1XR1 is implicated in opioid addiction through its potential involvement in the Wnt signaling pathway. Some mutations can amplify the activation of the Wnt pathway mediated by TBL1XR1\cite{nishi2017novo}. While there are indications of its role, a deeper exploration is essential to fully demonstrate TBL1XR1's significance in the context of opioid addiction.
    
    \subsection{Cocaine addiction-related key genes}
    To unravel the intricate interplay between our pinpointed key genes and cocaine addiction, we conducted a detailed literature validation for a comprehensive understanding.
    
    The immediate early gene FOS (c-Fos) has been a significant player in the context of cocaine addiction, as underscored by numerous studies exploring its intricate involvement in cocaine-induced neuroplasticity and behavioral responses\cite{hiroi1997fos, zhang2006c, mcclung2003regulation, dhonnchadha2012changes, xu2008c}. Zhang et al. investigated the influence of repeated cocaine administration on the expression of various molecular markers, particularly in Fos-deficient brain environments\cite{zhang2006c}. The result showed that Fos's absence in the brain results in altered expression levels of several transcription factors, neurotransmitter receptors, and intracellular signaling molecules ---all of which are induced by repeated cocaine exposure. This observation suggests that Fos is instrumental in acquiring cocaine-induced persistent changes. Xu demonstrated that the dendritic reorganization of medium spiny neurons, a hallmark of cocaine-induced neuroadaptations, was notably attenuated in Fos-mutant brains\cite{xu2008c}. These structural changes, or the lack thereof, were not merely restricted to the molecular domain; they manifested behaviorally as well. Mice with mutant FOS genes exhibited a marked reduction in behavioral sensitization, a phenomenon characterized by an escalating response to repeated cocaine administration. 
    
    Interleukin-6 (IL-6) has been suggested to have some connections with cocaine-induced behaviors. Mai et al.'s study on IL-6 knockout mice indicated a potential protective effect against cocaine-induced reactions, hinting at a role of the JAK2/STAT3 and PACAP signaling pathways\cite{mai20186}. Additionally, observed changes in serum levels of IL-6 among cocaine users hint at a peripheral response to the drug\cite{moreira2016cocaine}. However, the overall evidence linking IL-6 to cocaine effects is still limited, and more research is needed to establish a direct relationship.  
    
    Research suggests that SYT1 may influence cognitive performance from cocaine addiction. Silva et al. investigated the link between SYT1-rs2251214 and susceptibility as well as severity (as gauged by the addiction severity index) of CUD in a study involving 315 smoked cocaine addicts and 769 non-addicts\cite{da2019association}. Their findings highlighted a significant association between SYT1-rs2251214 and CUD vulnerability. Additionally, Viola et al. identified a correlation between cognitive performance and SYT1-rs2251214 in women diagnosed with cocaine use disorder\cite{viola2019association}. 
    
    Research has increasingly highlighted the significant role of $\alpha$-synuclein (SNCA) in various forms of substance addiction, extending beyond cocaine. SNCA is known for its critical involvement in dopaminergic transmission, a pathway often implicated in addictive behaviors\cite{sidhu2004role}. Qin et al. observed that chronic cocaine abuse leads to an upregulation of SNCA expression in the human striatum\cite{qin2005cocaine}. This was evidenced by immunoblot analysis in the ventral putamen, revealing elevated SNCA protein levels in striatal synaptosomes of cocaine users compared to age-matched drug-free controls. Additionally, a study by Foroud et al. suggests a link between SNCA variations and alcohol craving\cite{foroud2007association}. Although alcohol craving is a common feature of alcohol dependence, it is not universally present. Their findings indicate that genetic variation in SNCA could contribute to these craving behaviors, underscoring the gene's broader relevance in substance addiction.
    
    \subsection{Integrated Analysis of Opioid and Cocaine Addiction}
    Extensive literature validation underscores a significant association between GABRB3, PTPRN2, and GLS genes with drug addiction. The GABRB3 gene encodes the $\beta$ 3 subunit of the GABAA receptor. Chen et al. highlighted the potential of GABRB3 in heroin dependence, suggesting that its elevated expression may play a pivotal role in the disorder's pathogenesis \cite{chen2014association}. Furthermore, several studies have emphasized the role of GABRB3 in alcoholism. Noble et al. demonstrated that both DRD2 and GABRB3 variants heightened the risk for alcoholism \cite{noble1998d2}. Similarly, findings by Song et al. linked paternal transmission of GABRB3 to alcoholism \cite{song2003association}, while Young and colleagues associated both DRD2 and GABRB3 with alcohol-related expectations \cite{young2004alcohol}. PTPRN2 has emerged as a significant gene in substance dependence and cognitive behavior \cite{khan2023genetic}. Linkage analyses have identified its significance in the comorbidity of cocaine dependence and major depressive episodes in humans\cite{yang2011genomewide}. Further, genome-wide association studies have linked PTPRN2 to cognitive performance, risk-taking behaviors, and smoking initiation \cite{buniello2019nhgri}. Complementing the human findings, studies on PTPRN2 knockout mice have reported reduced concentrations of crucial neurotransmitters such as dopamine, norepinephrine, and serotonin in the brain \cite{nishimura2009disturbances}. Glutaminase (GLS) is an enzyme responsible for the conversion of the amino acid glutamine into glutamate\cite{porter2002complexity}. As the predominant excitatory neurotransmitter in the central nervous system, glutamate plays pivotal roles in the behavioral effects elicited by psychostimulant drugs\cite{marquez2017glutamate}. Over the past two decades, a combination of fundamental neuroscience research and preclinical studies using animal models has underscored the centrality of glutamate transmission in drug reward mechanisms, reinforcement, and the propensity for relapse\cite{d2015glutamatergic}. Given that glutaminases are the primary producers of glutamate in the brain, their potential significance in the realm of drug addiction becomes apparent\cite{marquez2017glutamate}. Glutamate's effects are mediated through two primary receptor types: ionotropic and metabotropic\cite{vikelis2007role}. The former are ion channels that facilitate ion movement upon glutamate activation, whereas the latter, being G-protein coupled receptors, trigger intricate signal transduction pathways when bound to glutamate. Notably, among these, NMDAR (an ionotropic receptor) and mGluR5 (a metabotropic receptor) are crucial in modulating neuronal excitability. Multiple studies have illuminated their intimate association with substance addiction\cite{bryant2006nmda, corbett2023mglu5, glass2011opioid, kato2006implication, kato2007role, mihov2016negative, salling2021negative}. As for IL1B, acute cocaine exposure has been linked to increased IL1B levels in specific brain regions, as shown by Cearley et al.'s findings in the cortex and nucleus accumbens\cite{cearley2011acute}. Montesinos et al. further observed that cocaine-induced changes in CX3CL1 concentrations are related to IL1B levels, suggesting activation of shared inflammatory pathways in the hippocampus\cite{montesinos2020cocaine}. The significance of IL-1B in addiction is further underscored by the findings of Liang et al., who explored its genetic aspects in relation to alcohol dependence\cite{liu2009association}. Their study revealed that polymorphisms in IL-1B are associated with a modified risk profile for alcohol dependence. Intriguingly, the particular single nucleotide polymorphisms (SNPs) at positions -511 and -31 are found with greater frequency in opioid-dependent populations. Although the association is described as weak, it nonetheless suggests a potential genetic predisposition linked to an increased risk of opioid dependence. This aligns with our identification of IL1B as a key gene in addiction, revealing its cross-substance implications and reinforcing its potential role as a biomarker or therapeutic target.
    
    \section{Topological objects}
    \subsection{Simplicial complex and chain complex}
    
    The PPI network is conceptualized as a graph where nodes correspond to proteins, and edges denote the interactions between pairs of proteins. Extending beyond the graph structure, we employ the concept of a simplicial complex to allow high-order interactions in the network,  
    encapsulating  a more comprehensive range of shapes and facilitating the inclusion of high-dimensional topological relationships. A simplicial complex is an aggregate of simplices, which are the fundamental building blocks that span various dimensions. Specifically, a $q$-simplex (denoted as $\sigma_q$) can be defined as the convex combination of $(q + 1)$ vertices $(v_0, v_1, v_2, ..., v_q)$ that are affinely independent, expressed as:
    \begin{equation}\label{eq:12}
        \sigma_q := [ v_0, v_1, ..., v_q].
    \end{equation}
    In the realm of Euclidean geometry, simplices correspond to familiar shapes: a 0-simplex represents a point, a 1-simplex forms a line segment, a 2-simplex constitutes a triangle, and a 3-simplex is akin to a tetrahedron. For any given set of $(q + 1)$ points, every non-empty subset can be the vertices of a subsimplex, considered to be a face of a $q$-simplex, denoted as $\sigma^m\subset \sigma_q$. The structure known as a simplicial complex, denoted by $K$, is defined as a finite set of simplices that adhere to two critical conditions:
    \begin{enumerate}
        \item[1)] Any face of a simplex within $K$ must also be included in $K$;
        \item[2)] The nonempty intersection of any two simplices is a face of both simplices. 
    \end{enumerate}
    
    The relationship between simplices within a complex can be characterized by their adjacency, a concept extended from the combinatorial graph theory. In a graph, the degree of a vertex, denoted as deg$(v)$, is the tally of its adjacent edges. This principle becomes more complex when applied to $q$-simplices, which are adjacent to both $(q - 1)$-simplices and $(q + 1)$-simplices. To navigate this complexity, it is necessary to distinguish between upper and lower adjacencies for defining the degree of a $q$-simplex when $q>0$. Two $q$-simplices, $\sigma^1_q$ and $\sigma^2_q$, within a complex $K$ are considered upper adjacent $\sigma^1_q \stackrel{U}\sim \sigma^2_q$ if they both constitute faces of the same $(q + 1)$-simplex. Conversely, they are lower adjacent $\sigma^1_q \stackrel{L}\sim \sigma^2_q$ if they share a common $(q - 1)$-simplex. The upper degree, deg$_U(\sigma_q)$, is the count of $(q + 1)$-simplices of which the $q$-simplex is a face, while the lower degree, deg$_L(\sigma_q)$, is the count of its $(q - 1)$-simplex faces, which invariably equals $(q + 1)$. Consequently, for $q$-simplices where $q > 0$, the degree is the sum of the upper and lower degrees:
    \begin{equation}
        \text{deg}(\sigma_q) = \text{deg}_L(\sigma_q) + \text{deg}_U(\sigma_q).
    \end{equation}
    
    The concept of orientation in a simplex is pivotal and is dictated by the sequence of its vertices, excluding the 0-simplex which has no orientation. For a $q$-simplex $\sigma_q$, two given vertex orderings are considered to be similarly oriented if they can be interchanged by an even permutation, hence both configurations represent an orientation of $\sigma_q$. If an odd permutation is required, the orderings are dissimilarly oriented. Consequently, an oriented$q$-simplex is the pairing of a simplex $\sigma_q$ with a designated orientation. When every simplex within a simplicial complex $K$ is assigned an orientation, we refer to $K$ as an oriented simplicial complex. 
    
    In topological, geometric, and algebraic studies, the framework of chain complexes is essential. Consider a simplicial complex $K$ with a maximum dimension of $q$. Within this context, a $q$-chain is constructed as a formal summation of the $q$-simplices in $K$, utilizing coefficients drawn from the $\mathbb{Z}_2$ field. The collection of all $q$-chains, under the addition defined by $\mathbb{Z}_2$, forms a group known as the chain group, symbolized by $C_q(K)$. To create connections between chain groups across various dimensions, the boundary operator for $q$-chains, denoted $\partial_q$, serves as a mapping function from $C_q(K)$ to $C_{q-1}(K)$. This operator transforms a $q$-chain, represented as a linear combination of $q$-simplices, into the corresponding combination of their $(q - 1)$-dimensional boundaries. For a $q$-simplex spanned by vertices $[v_0, v_1,\cdots,v_q]$, symbolized as $\sigma_q$, the action of $\partial_q$ is mathematically articulated as: 
    \begin{equation}
        \partial_q \sigma_q = \sum_{i=0}^{q}(-1)^i[v_0, \cdots, \hat{v_i},\cdots,v_q],
    \end{equation}
    where $\hat{v_i}$ indicates the $(q - 1)$-simplex formed by excluding the vertex $v_i$ from $\sigma_q$. A $q$-chain that yields a zero boundary upon the application of $\partial_q$ is termed a $q$-cycle.
    
    The structure of a chain complex is characterized by a sequence of chain groups linked by boundary operators. This sequence forms a continuous cascade from higher-dimensional chains down to zero-dimensional chains, which is mathematically depicted as:
    \begin{equation}
        \cdots \stackrel{\partial_{q+2}}\longrightarrow C_{q+1}(K) \stackrel{\partial_{q+1}}\longrightarrow C_{q}(K) \stackrel{\partial_{q}}\longrightarrow C_{q-1}(K)\stackrel{\partial_{q-1}} \longrightarrow \cdots
    \end{equation}
    In this expression, $C_{q}(K)$ represents the chain group composed of $q$-chains in the simplicial complex $K$, while $\partial_{q}$ denotes the boundary operator mapping $q$-chains to $(q-1)$-chains. The chain complex terminates with the null set, indicating that the boundary of a 0-chain is inherently zero since there are no negative-dimensional chains. Each boundary operator connects adjacent levels of the complex, ensuring that the boundary of a boundary within this sequence is always zero, a foundational concept in homology theory.
    
    \subsection{$q$-combinatorial Laplacian}
    
    In the field of algebraic topology, the boundary operators within a simplicial complex $K$ facilitate the construction of a bridge between different dimensions of chains. The matrix representation of the $q$-boundary operator $\partial_q: C_q(K) \longrightarrow C_{q-1}(K)$ is denoted as $\mathcal{B}_q$. The dimensions of this matrix—rows corresponding to the number of $(q - 1)$-simplices and columns to the number of $q$-simplices—reflect the structure of $K$. The adjoint of $\partial_q$, symbolized as $\partial_q^{\ast}: C_{q-1}(K) \longrightarrow C_q(K)$, operates in the reverse direction of the boundary operator, and its matrix representation is the transpose of $\mathcal{B}_q$, labeled $\mathcal{B}_q^T$. Diving deeper into topological invariants, the $q$-combinatorial Laplacian is a linear transformation $\Delta_q: C_q(K) \longrightarrow C_q(K)$ defined as: 
    \begin{equation}\label{equ:laplacian operator}
        \Delta_q := \partial_{q+1} \partial_{q+1}^{\ast} + \partial_{q}^{\ast} \partial_{q}.
    \end{equation}
    Its matrix form, $\mathcal{L}_q$, emerges as a summation of the product of boundary matrices and their transposes:
    \begin{equation}\label{equ:combinatorial Laplacian}
        \mathcal{L}_q = \mathcal{B}_{q+1}\mathcal{B}_{q+1}^{T} + \mathcal{B}_q^T \mathcal{B}_q.
    \end{equation}
    Specifically, for $q = 0$, the combinatorial Laplacian, also known as the graph Laplacian, simplifies due to $\partial_0$ being a zero map, resulting in:
    \begin{equation}
        \mathcal{L}_0 = \mathcal{B}_{1}\mathcal{B}_{1}^{T}.
    \end{equation}
    The matrix elements of $\mathcal{L}_q$ are defined by the relationships between simplices and their degrees of connectivity:
    \begin{equation}
        (\mathcal{L}_q)_{ij} = 
        \begin{cases}\label{equ:combine}
            \text{deg}(\sigma^i_{q}) + q + 1, & \mbox{if $i=j$.} \\
            1,                 & \mbox{if $i\neq j$, $\sigma^i_q \stackrel{U}\nsim \sigma^j_q$ and $\sigma^i_q \stackrel{L}\sim \sigma^j_q$ with similar orientation.}   \\
            -1,                & \mbox{if $i\neq j$, $\sigma^i_q \stackrel{U}\nsim \sigma^j_q$ and $\sigma^i_q \stackrel{L}\sim \sigma^j_q$ with dissimilar orientation.} \\
            0,                 & \mbox{if $i\neq j$ and either , $\sigma^i_q \stackrel{U}\sim \sigma^j_q$ or $\sigma^i_q \stackrel{L}\nsim \sigma^j_q$.}
        \end{cases} \\
    \end{equation}
    For the graph Laplacian, when $q = 0$, the matrix $\mathcal{L}_0$ is simplified as:
    \begin{equation}
        (\mathcal{L}_0)_{ij} = 
        \begin{cases}\label{equ:L0}
            \text{deg}(\sigma^i_{0}),   & \mbox{if $i=j$.} \\
            -1,                         & \mbox{if $\sigma^i_0 \stackrel{U}\sim \sigma^j_0$.} \\
            0,                          & \mbox{otherwise.}
        \end{cases}
    \end{equation}
    The $q$-combinatorial Laplacian matrix is a pivotal element in understanding the intrinsic topological features of a simplicial complex. Due to its symmetric and positive semi-definite nature, it is guaranteed that all eigenvalues of the matrix are real and non-negative. These spectral properties play a central role in revealing the topological invariants of the underlying space.
    According to the combinatorial Hodge theorem, the Betti numbers, which serve as topological invariants representing the number of $q$-dimensional holes in a space, can be computed from the spectrum of the Laplacian. The multiplicity of the zero eigenvalue, also known as the nullity of the Laplacian matrix $\mathcal{L}_q$, corresponds to the $q$-th Betti number ${\beta_q}$:
    \begin{equation}\label{equ:betti}
        \beta_q = \text{dim}(\mathcal{L}_q(K)) - \text{rank}(\mathcal{L}_q(K)) = \text{nullity}(\mathcal{L}_q(K)) = \# ~{\rm of ~ zero ~eigenvalues~ of}~   \mathcal{L}_q(K).
    \end{equation}
    These Betti numbers encapsulate crucial topological information. ${\beta_0}$ indicates the number of connected components within the complex. ${\beta_1}$ represents the number of one-dimensional or ``circular`` holes, often conceptualized as tunnels or loops. ${\beta_2}$ corresponds to the number of two-dimensional voids or ``cavities", akin to the hollow spaces enclosed by surfaces. 
    
    \section{Datasets for machine learning}
    
\begin{figure}[H]
        \centering
        \includegraphics[width = 1\textwidth]{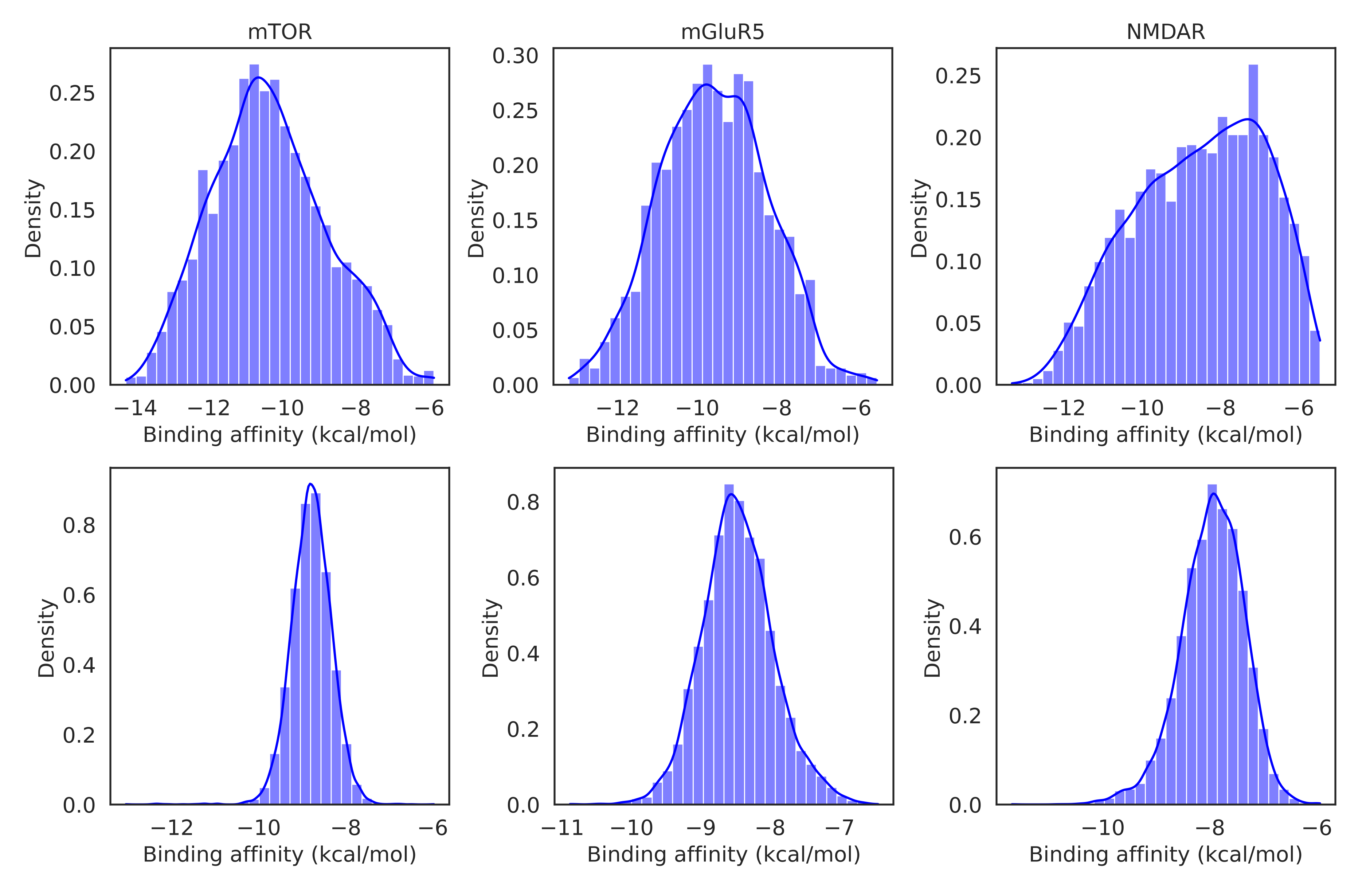}
        \caption{Binding Affinity Distributions in Datasets: The top panel displays the distribution of binding affinity values extracted from the ChEMBL dataset. The bottom panel illustrates the distribution of predicted binding affinities for drugs in DrugBank.}
        \label{fig:BA_dis}
    \end{figure}

    In our study, we developed predictive models for binding affinity using inhibitor datasets sourced from the ChEMBL database (Table \ref{tab: dataset}). These models were then applied to estimate the binding affinity of various drugs listed in DrugBank towards three specific targets. The compounds within these datasets were represented using SMILES strings, and their corresponding binding affinities, measured in kcal/mol, were used as the target labels for model training. The distribution of these binding affinity labels across each dataset is depicted in Figure \ref{fig:BA_dis}.
    
    \begin{table}[H] 
        \centering
        \caption{The summary of datasets used in this study.}
        \label{tab: dataset}
        {
        \begin{tabular}{ cccc } 
    \toprule
    Dataset &Protein name &Dataset size &Binding affinity range (kcal/mol)\\
    \midrule
    mTOR &Mammalian target of rapamycin &4392 &[-5.86, -14.25]\\
    mGluR5 &The metabotropic glutamate receptor subtype 5 &1777 &[-5.47, -13.23]\\
    NMDAR &N-methyl-D-aspartate receptor &2342 &[-5.45, -13.32]\\
        \bottomrule
        \end{tabular} 
        }
    \end{table}
    
    We constructed predictive models for binding affinity using inhibitor datasets obtained from the ChEMBL database. We then utilize these models to predict the binding affinity of drugs in DrugBank to these three targets. The compounds in these datasets are represented by SMILES strings, and their binding affinities (kcal/mol) serve as labels. The distribution of labels for each dataset is illustrated in Figure S1.
    
    \section{Evaluation metrics}
    To assess the performance of our regression models, we employed two key metrics: the Pearson Correlation Coefficient (PCC) and the Root Mean Squared Error (RMSE). The definitions and formulas for these metrics are as follows:
    The Pearson Correlation Coefficient is defined for two vectors $x = (x_1, x_2, \dots, x_n)$ and $y = (y_1, y_2, \dots, y_n)$ as:
    \begin{equation}\label{equ:pcc}
        R = \text{PCC}(x,y) = \frac{\sum (x_i - \overline{x})(y_i - \overline{y})}
    {\sqrt{\sum (x_i - \overline{x})^2 \sum (y_i - \overline{y})^2}}
    \end{equation}
    where $\overline{x}$ and $\overline{y}$ represent the mean of vectors $x$ and $y$, respectively.
    The Root Mean Squared Error (RMSE) is calculated as:
    \begin{equation}
        \text{RMSE} = \sqrt{\frac{1}{n} \sum_{i=1}^{n} (y_i - \hat{y}_i)^2},
    \end{equation}
    where $y_i$ is the true value and $\hat{y}_i$ is the predicted value for the $i$-th sample.
    
    \section{Gene significance ranking}
    We employed topological differentiation to assess the significance of each gene within the PPI network through a multiscale analysis approach. Utilizing networks set at four different thresholds allowed us to conduct a comprehensive multi-resolution analysis. The key genes were identified by intersecting the top 25 genes from each of these four threshold-defined networks, ensuring a robust and consistent selection of significant genes (Table \ref{tab: opioid_ranking}, \ref{tab: cocaine_ranking}).
    
    \begin{table}[H] 
        \centering
        \caption{Top 25 Key Genes in the Opioid Addiction-Related DEG PPI Network. The key genes were identified as the intersection of the top-ranking genes across four networks, each defined by different threshold levels.}
        \label{tab: opioid_ranking}
        {
        \begin{tabular}{ ccccc } 
    \toprule
    Ranking &Threshold 0.15 &Threshold 0.4 &Threshold 0.7 &Threshold 0.9\\
    \midrule
    1 &YWHAZ &PML &\textbf{BUB1} &PML\\
    2 &MAP2K1 &YWHAZ &PML &STAT6\\
    3 &NTRK2 &EIF5B &EIF5B &\textbf{BUB1}\\
    4 &PGK1 &PECAM1 &STAT6 &EIF5B\\
    5 &UBE2N &\textbf{BUB1} &NCOA1 &NCOA1\\
    6 &\textbf{BUB1} &\textbf{ADCY9} &\textbf{RHEB} &LRP6\\
    7 &TPM2 &NTRK2 &SRD5A1 &TP63\\
    8 &TP63 &\textbf{TBL1XR1} &\textbf{APC} &\textbf{RHEB}\\
    9 &IL4 &MCM5 &\textbf{ADCY9} &SRD5A1\\
    10 &PPP3CA &NCOA1 &MCM5 &SORT1\\
    11 &MYH11 &RPL35A &YWHAZ &WNT8B\\
    12 &REL &MKI67 &\textbf{TBL1XR1} &ARNT\\
    13 &SOX5 &IL4 &PRKAR1A &CDKN1C\\
    14 &PECAM1 &SIL1 &TCAP &HSD17B6\\
    15 &PIDD1 &RAPGEF3 &NPAS2 &\textbf{ADCY9}\\
    16 &\textbf{APC} &\textbf{APC} &TP63 &\textbf{APC}\\
    17 &WDFY3 &DOHH &DDX28 &\textbf{TBL1XR1}\\
    18 &HMGB1 &HBEGF &SORT1 &NPAS2\\
    19 &\textbf{TBL1XR1} &STAT6 &CDKN1C &RBM25\\
    20 &RAB40B &NPAS2 &HSD17B6 &DOHH\\
    21 &\textbf{ADCY9} &\textbf{RHEB} &RBM25 &EIF5A2\\
    22 &MKI67 &PRKAR1A &UBE2N &LUC7L3\\
    23 &RET &EIF5A2 &RET &IL4\\
    24 &\textbf{RHEB} &PGK1 &ARNT &NTRK2\\
    25 &PRDX3 &NOC4L &H2AFV &EIF1AX\\
        \bottomrule
        \end{tabular} 
        }
    \end{table}
    
    \begin{table}[H] 
        \centering
        \caption{Top 25 Key Genes in the Cocaine Addiction-Related DEG PPI Network. The key genes were identified as the intersection of the top-ranking genes across four networks, each defined by different threshold levels.}
        \label{tab: cocaine_ranking}
        {
        \begin{tabular}{ ccccc } 
    \toprule
    Ranking &Threshold 0.15 &Threshold 0.4 &Threshold 0.7 &Threshold 0.9\\
    \midrule
    1 &\textbf{SNAP25} &\textbf{SNAP25} &\textbf{JUN} &\textbf{JUN}\\
    2 &\textbf{JUN} &\textbf{JUN} &\textbf{IL6} &\textbf{IL6}\\
    3 &\textbf{IL6} &\textbf{IL6} &\textbf{SNAP25} &VAMP2\\
    4 &GRIN1 &BDNF &VAMP2 &\textbf{SYT1}\\
    5 &\textbf{FOS} &GRIN1 &\textbf{SYT1} &\textbf{FOS}\\
    6 &YWHAH &SYP &FGF2 &CD44\\
    7 &BDNF &\textbf{SNCA} &BDNF &FGF2\\
    8 &\textbf{SYT1} &IL1B &CD44 &\textbf{SNAP25}\\
    9 &GAD2 &CD44 &\textbf{SNCA} &EGR1\\
    10 &\textbf{DNM1} &SYN1 &\textbf{FOS} &\textbf{SNCA}\\
    11 &MAST1 &SOX2 &EGR1 &IRF1\\
    12 &SYP &GAP43 &CDK5 &IL1B\\
    13 &\textbf{SNCA} &CALM3 &PAK1 &CACNG2\\
    14 &ENO2 &FGF2 &SLC32A1 &\textbf{DNM1}\\
    15 &IL1B &\textbf{FOS} &VCAM1 &CDK5\\
    16 &STMN2 &\textbf{SYT1} &GRIN1 &PAK1\\
    17 &SNCB &SLC32A1 &GAD2 &CCL2\\
    18 &NTRK2 &VAMP2 &STXBP1 &JUNB\\
    19 &SYN1 &GAD2 &\textbf{DNM1} &SLC6A3\\
    20 &GAP43 &EDN1 &BAG3 &BAG3\\
    21 &ATP2B2 &\textbf{DNM1} &CACNG2 &CXCL10\\
    22 &GABRG2 &STXBP1 &CALM3 &STXBP1\\
    23 &CALM3 &KCNQ2 &DNAJB1 &CDKN1A\\
    24 &KCNQ2 &TH &SYP &HSPA1A\\
    25 &SLC32A1 &SNCB &HSPA1A &NCAN\\
        \bottomrule
        \end{tabular} 
        }
    \end{table}
    

\end{appendices}
\bibliographystyle{unsrt}
    \bibliography{main}

\end{document}